\documentclass[10pt]{revtex4-1}    
\usepackage{amssymb}
\usepackage{latexsym}
\usepackage{epsfig}
\begin{document}

\title{Quintessence reconstruction of interacting HDE in a non-flat universe}

\author{A. Sheykhi$^{1,2}$\footnote{asheykhi@shirazu.ac.ir},
H. Alavirad$^{3}$\footnote{hamzeh.alavirad@partner.kit.edu}
A. Bagheri$^{1}$ and
E. Ebrahimi $^{4}$ \footnote{eebrahimi@uk.ac.ir}}

\address{$^1$ Physics Department and Biruni Observatory, College of
Sciences, Shiraz University, Shiraz 71454, Iran\\
         $^2$ Research Institute for Astronomy and Astrophysics of Maragha
         (RIAAM), P.O. Box 55134-441, Maragha, Iran\\
         $^3$ Institute for Theoretical Physics, Karlsruhe Institute of Technology, 76128 Karlsruhe, Germany\\
         $^4$  Department of Physics, Shahid Bahonar University, P.O. Box 76175, Kerman, Iran}

\begin{abstract}

In this paper we consider quintessence reconstruction of
interacting holographic dark energy in a non-flat background. As
system's IR cutoff we choose the radius of the event horizon
measured on the sphere of the horizon, defined as $L=ar(t)$. To
this end we construct a quintessence model by a real, single
scalar field. Evolution of the potential, $V(\phi)$, as well as
the dynamics of the scalar field, $\phi$, are obtained according
to the respective holographic dark energy. The reconstructed
potentials show a cosmological constant behavior for the present
time. We constrain the model parameters in a flat universe by
using the observational data, and applying the Monte Carlo Markov
chain simulation. We obtain the best fit values of the holographic
dark energy model and the interacting parameters as
$c=1.0576^{+0.3010+0.3052}_{-0.6632-0.6632}$ and
$\zeta=0.2433^{+0.6373+0.6373}_{-0.2251-0.2251}$, respectively.
From the data fitting results we also find that the model can
cross the phantom line in the present universe where the best fit
value of of the dark energy equation of state is  $w_D=-1.2429$.\\
\textit{keywords:} quintessence; holographic; dark energy; data fitting.
\end{abstract}
\maketitle

\section{Introduction\label{Int}}
A wide range of observational evidences support the present
acceleration of the Universe expansion. The first evidence for the
mentioned acceleration is the cosmological observations from Type
Ia supernovae (SN Ia) \cite{snia}. Subsequently such acceleration
was repeatedly confirmed by Cosmic Microwave Background (CMB)
anisotropies measured by the WMAP satellite \cite{cmb}, Large
Scale Structure \cite{lss}, weak lensing \cite{weaklen} and the
integrated Sach-Wolfe effect \cite{swolf}. Based on the Einstein's
theory of gravity, such an acceleration needs an exotic type of
matter with negative pressure, usually called dark energy (DE) in
the literatures. This new component consists more than $70\%$ of
the present energy content of the universe. The simplest
alternative which can explain the phase of acceleration is the so
called cosmological constant which originally was presented by
Einstein to build a static solution for the universe in the
context of general relativity. Although cosmological constant can
explain the acceleration of the universe but it suffers the ``fine
tuning" and ``coincidence" problems. Of interesting models of DE
are those which called scalar field models. A typical property of
these models is their time varying equation of state parameter
($w=\frac{P}{\rho}$) favored by cosmic observations
\cite{feng,alam,huterer}. A plenty of these models have been
presented in the literature which an incomplete list is
quintessence, tachyon, K-essence, agegraphic, ghost and so on (see
\cite{Cop,Li1,ASH} and references therein).

Among different candidate to DE, holographic dark energy (HDE) is
one which contains interesting features. This model is based on
the holographic principle which states that the entropy of a
system scales not with it's volume, but with it's surface area
\cite{Suss1} and it should be constrained by an infrared cutoff
\cite{Coh}. Applying such a principle to the DE issue and taking
the whole universe into account, then the vacuum energy related to
this holographic principle is viewed as DE, usually called HDE
\cite{Coh,hsu,li}. According to these statements the holographic
energy density can be written as \cite{Coh}
\begin{equation}\label{rhoEins}
\rho_{D}= \frac{3c^2M^2_p}{L^2},
\end{equation}
where $c^2$ is a numerical constant, $M_p^{2} =( 8\pi G)^{-1}$,
and $L$ is an infrared (IR) cutoff radius. It is worth mentioning
that the holographic principle does not determine the IR cutoff
and we have still freedom to choose $L$. Different choices for IR
cutoff parameter, $L$, have been proposed in the literature, among
them are, the particle horizon \cite{particle}, the future event
horizon \cite{future}, the Hubble horizon \cite{hsu2,pavon} and
the apparent horizon \cite{shey1}. Each of these choices solve
some features and lead new problems. For instance in the HDE model
with Hubble horizon, the fine tuning problem is solved and the
coincidence problem is also alleviated, however, the effective
equation of state for such vacuum energy is zero and the universe
is decelerating \cite{hsu2} unless the interaction is taken into
account \cite{pavon}. For a complete list of papers concerning HDE
one can refer to \cite{HDE} and references therein.

Nowadays, every model which can explain the acceleration of the
Universe expansion, and is consistent with observational
evidences, could be accepted as a DE candidate. Due to the lack of
observational evidences about DE models, many approaches are
presented to answer the puzzle of the unexpected acceleration of
the Universe. The number and variety of these models are so
increasing which we should classify them in any way. One main task
in this way is to find equivalent theories presented in different
frameworks, however, they seems to have distinct origins. One
valuable approach which recently has attracted a lot of attention
is to make scalar field dual of the DE models
\cite{XZ,Setare,tachADE,quinADE,ghost}. This interest in the
scalar field models of DE partly comes from the fact that scalar
fields naturally arise in particle physics including
supersymmetric field theories and string/M theory. Beside with
clarifying the status of the models in the literature maybe
sometimes we can use the corresponding scalar field dual of DE
model predicting new features and setting observational
constraints on the free parameters.

In this paper our aim is to establish a correspondence between the
HDE and quintessence model of DE in a non-flat universe. As
systems's IR cutoff we shall choose the radius of the event horizon
measured on the sphere of the horizon, defined as $L=ar(t)$.
Quintessence assumes a canonical scalar field $\phi$ and a self
interacting potential $V(\phi)$ minimally coupled to the other
component in the universe. Quintessence is described by the
Lagrangian of the form
\begin{equation}\label{qelag}
{\cal
L}=-\frac{1}{2}g^{\mu\nu}\partial_{\mu}\phi\partial_{\nu}\phi-V(\phi).
\end{equation}
The energy-momentum tensor of quintessence is
\begin{equation}\label{emtensqe}
    T_{\mu\nu}=\partial_{\mu}\phi\partial_{\nu}\phi-g_{\mu\nu}
    \left[\frac{1}{2}g^{\alpha\beta}\partial_{\alpha}\phi\partial_{\beta}\phi+V(\phi)\right].
\end{equation}
In quintessence model we choose a convenient potential $V(\phi)$
to obtain desirable result in agreement with observations. Hence,
our goal in this paper is to reconstruct the potential $V(\phi)$
corresponds to the HDE and investigate the evolution of different
parameters in the model. Our work differs from Ref. \cite{XZ} in
that we consider the interacting HDE model in a non-flat universe,
while the author of \cite{XZ} studied the non-interacting case in
a flat universe. It also differs from Refs. \cite{sheyHDE,GO}, in
that we take $L=ar(t)$ as system's IR cutoff not the Hubble radius
$L=H^{-1}$ proposed in \cite{sheyHDE}, nor the Ricci scalar like
cutoff, $L^{-2}=\alpha H^2+\beta \dot{H}$, introduced in
\cite{GO}.

This paper is organized as follows. In section \ref{HDE}, we
reconstruct the non-interacting holographic quintessence model
with $L=ar(t)$ as IR cutoff. In section \ref{IntHDE}, we extend
our study to the case where there is an interaction between DE and
dark matter. In order to check the viability of the model, in
section \ref{fitting}, we constrain the holographic interacting
quintessence model by using the cosmological data. We summarize
our results in section \ref{CONC}.
\section{Quintessence reconstruction of HDE  \label{HDE}}
Consider the non-flat Friedmann-Robertson-Walker (FRW) universe
which is described by the line element
\begin{eqnarray}
 ds^2=dt^2-a^2(t)\left(\frac{dr^2}{1-kr^2}+r^2d\Omega^2\right),\label{metric}
 \end{eqnarray}
where $a(t)$ is the scale factor, and $k$ is the curvature
parameter with $k = -1, 0, 1$ corresponding to open, flat, and
closed universes, respectively. The first Friedmann equation is
\begin{eqnarray}\label{Fried}
H^2+\frac{k}{a^2}=\frac{1}{3M_p^2} \left( \rho_m+\rho_D \right).
\end{eqnarray}
We introduce, as usual, the fractional energy densities such as
\begin{eqnarray}\label{Omega}
\Omega_m=\frac{\rho_m}{3M_p^2H^2}, \hspace{0.5cm}
\Omega_D=\frac{\rho_D}{3M_p^2H^2},\hspace{0.5cm}
\Omega_k=\frac{k}{H^2 a^2},
\end{eqnarray}
thus, the Friedmann equation can be written
\begin{eqnarray}\label{Fried2}
\Omega_m+\Omega_D=1+\Omega_k.
\end{eqnarray}
We shall assume the quintessence scalar field model of DE is the
effective underlying theory. The energy density and pressure for
the quintessence scalar field are given by \cite{Cop}
\begin{eqnarray}\label{rhophi}
\rho_\phi=\frac{1}{2}\dot{\phi}^2+V(\phi),\\
p_\phi=\frac{1}{2}\dot{\phi}^2-V(\phi). \label{pphi}
\end{eqnarray}
Thus the potential and the kinetic energy term can be written as
\begin{eqnarray}\label{vphi}
&&V(\phi)=\frac{1-w_D}{2}\rho_{\phi},\\
&&\dot{\phi}^2=(1+w_D)\rho_\phi. \label{ddotphi}
\end{eqnarray}
Next we implement the HDE model with quintessence field. The
holographic energy density has the form (\ref{rhoEins}), where the
radius $L$ in a nonflat universe is chosen as
\begin{equation}\label{L}
L=ar(t),
\end{equation}
and the function $r(t)$ can be obtained from the following
relation
\begin{equation}
 \int_{0}^{r(t)}{\frac{dr}{\sqrt{1-kr^2}}}=\int_{0}^{\infty}{\frac{dt}{a}}=\frac{R_h}{a}.
\end{equation}
It is important to note that in the non-flat universe the
characteristic length which plays the role of the IR-cutoff is the
radius $L$ of the event horizon measured on the sphere of the
horizon and not the radial size $R_h$ of the horizon. Solving the
above equation for general case of the non-flat FRW universe, we
have
\begin{equation}
r(t)=\frac{1}{\sqrt{k}}\sin y,\label{rt}
\end{equation}
where $y=\sqrt{k} R_h/a$. For latter convenience we rewrite the
second  Eq. (\ref{Omega}) in the form
\begin{eqnarray}
HL=\frac{c}{\sqrt{\Omega_D}}. \label{HL}
\end{eqnarray}
Taking derivative with respect to the cosmic time $t$ from Eq.
(\ref{L}) and using Eqs. (\ref{rt}) and (\ref{HL}) we obtain
\begin{eqnarray}
\dot{L}=HL+a\dot{r}(t)=\frac{c}{\sqrt{\Omega_D}}-\cos y.
\label{Ldot}
\end{eqnarray}
Consider the FRW universe filled with DE and dust (dark matter)
which evolves according to their conservation laws
\begin{eqnarray}
&&\dot{\rho}_D+3H\rho_D(1+w_D)=0,\label{consq1}\\
&&\dot{\rho}_m+3H\rho_m=0, \label{consm1}
\end{eqnarray}
where $w_D$ is the equation of state parameter of DE. Taking the
derivative  of Eq. (\ref{rhoEins}) with respect to time and using
Eq. (\ref{Ldot}) we find
\begin{eqnarray}
\dot{\rho}_D=-2H\rho_D\left(1-\frac{\sqrt{\Omega_D}}{c}\cos
y\right)\label{rhodot}.
\end{eqnarray}
Inserting this equation in conservation law (\ref{consq1}), we
obtain the equation of state parameter
\begin{eqnarray}
w_D=-\frac{1}{3}-\frac{2\sqrt{\Omega_D}}{3c}\cos y\label{wD1}.
\end{eqnarray}
Differentiating Eq. (\ref{HL}) and using relation
${\dot{\Omega}_D}={\Omega'_D}H$, we reach
\begin{eqnarray}\label{Omegaqn2}
{\Omega'_D}=\Omega_D\left(-2\frac{\dot{H}}{H^2}-2+\frac{2}{c
}\sqrt{\Omega_D}\cos y\right),
\end{eqnarray}
where the dot and the prime denote the derivative with respect to
the cosmic time and $x=\ln{a}$, respectively. Taking the derivative
of both side of the Friedman equation (\ref{Fried}) with respect to
the cosmic time, and using Eqs. (\ref{Fried2}), (\ref{HL}),
(\ref{consq1}) and (\ref{consm1}), it is easy to show that
\begin{eqnarray}\label{Hdot}
\frac{2\dot{H}}{H^2}=-3-\Omega_k+\Omega_D+\frac{2\Omega^{3/2}_D}{c}\cos
y.
\end{eqnarray}
Substituting this relation into Eq. (\ref{Omegaqn2}), we obtain the
equation of motion of HDE
\begin{eqnarray}\label{Omegaq3}
{\Omega'_D}&=&\Omega_D\left[(1-\Omega_D)\left(1+\frac{2}{c}\sqrt{\Omega_D}\cos
y\right)+\Omega_k\right].
\end{eqnarray}
We have plotted in Figs. 1 and 2 the evolutions of the $w_D$ and
$\Omega_D$ for the HDE with different parameter $c$. One can see
from Fig. 1 that increasing $c$ leads to a faster evolution of
$w_D$ toward more negative values, while a reverse behavior is
seen for $\Omega_D$ and increasing $c$ results a slower evolution
of $\Omega_D$.

Now we suggest a correspondence between the HDE and quintessence
scalar field namely, we identify $\rho_\phi$ with $\rho_D$. Using
relation $\rho_\phi=\rho_D={3M_p^2H^2}\Omega_D$ and Eq. (\ref{wD1})
we can rewrite the scalar potential and kinetic energy term as
\begin{figure*}[htp]
\begin{center}
\begin{tabular}{cc}
\epsfig{file=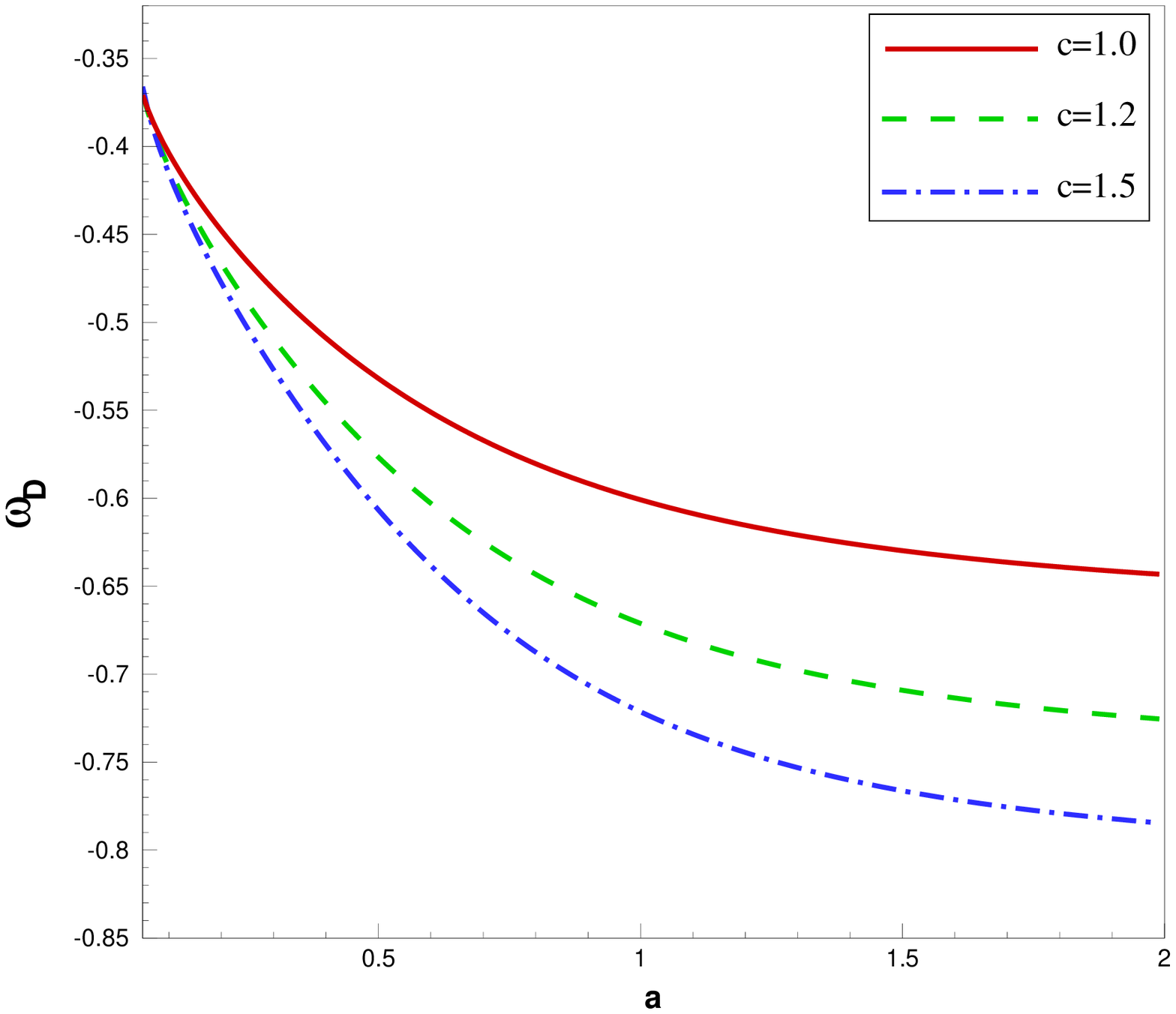, ,width=0.5\linewidth} & \epsfig{file=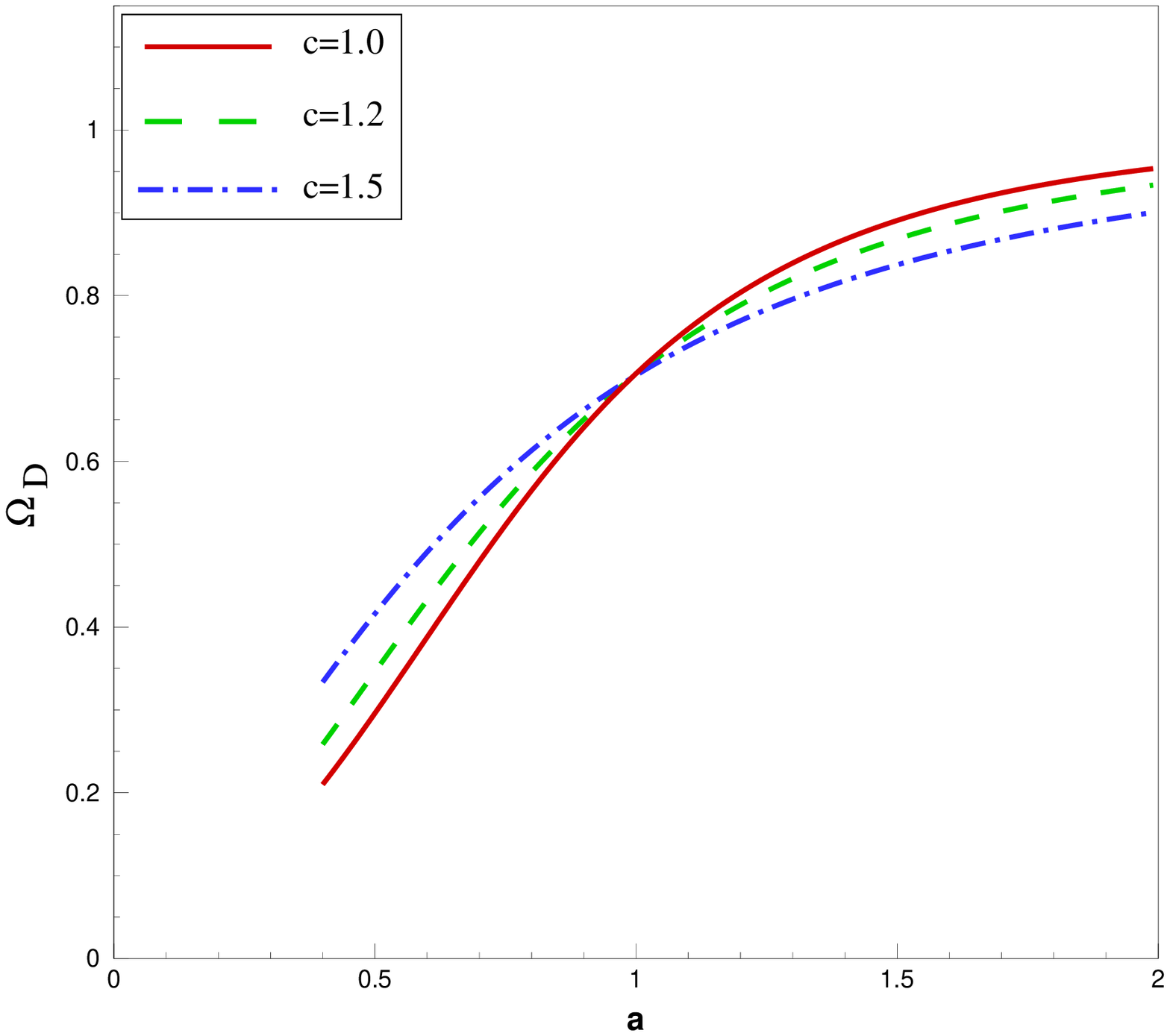, ,width=0.5\linewidth} \\
\end{tabular}
\caption{The evolution of $w_D$ (left) and $\Omega_{D}$ (right)
for HDE with different parameter $c$. Here we take
$\Omega_{D0}=0.72$ and $\Omega_{k}=0.01.$  } \label{fig1}
\end{center}
\end{figure*}
\begin{figure*}[htp]
\begin{center}
\begin{tabular}{cc}
\epsfig{file=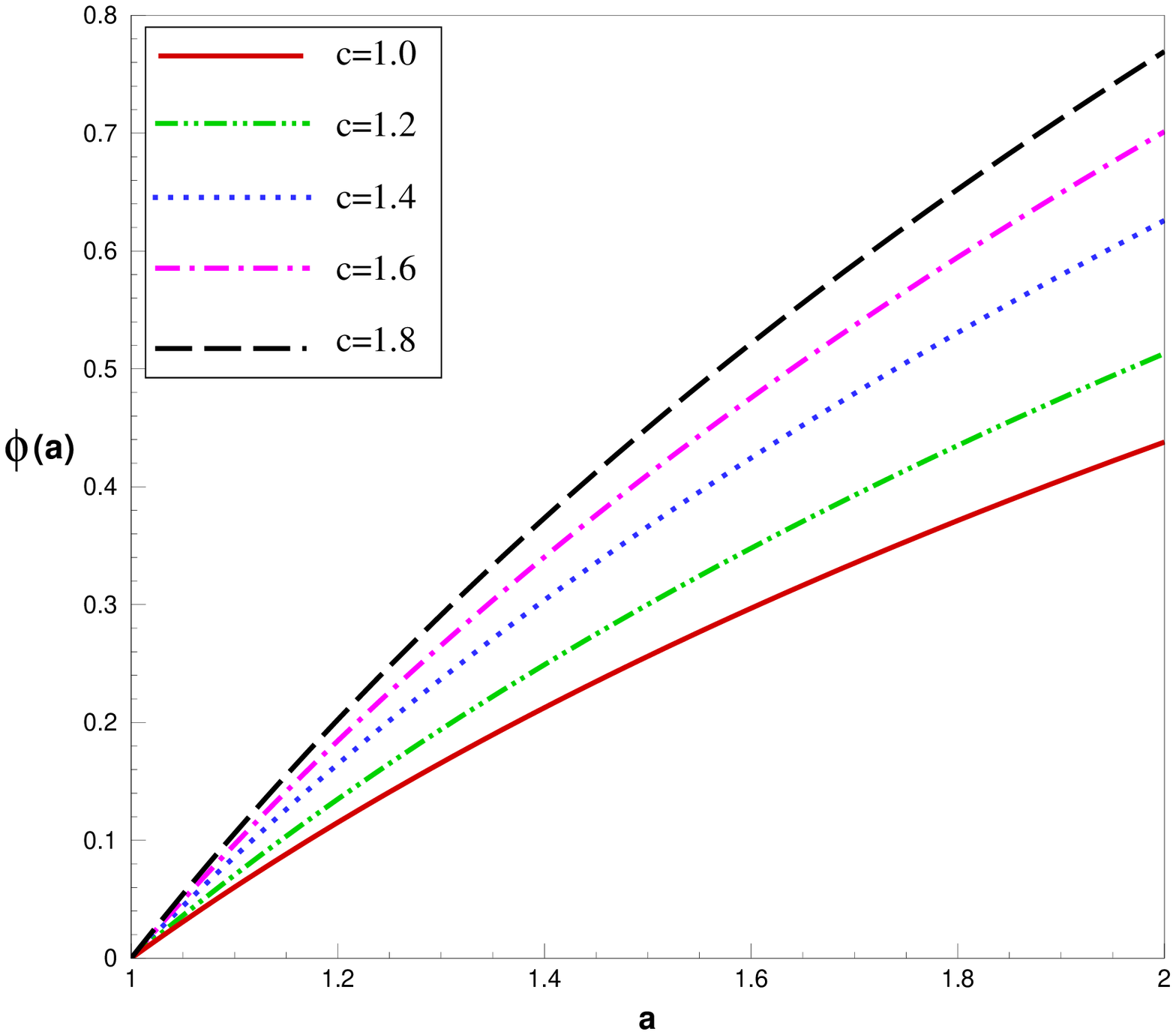, ,width=0.5\linewidth} & \epsfig{file=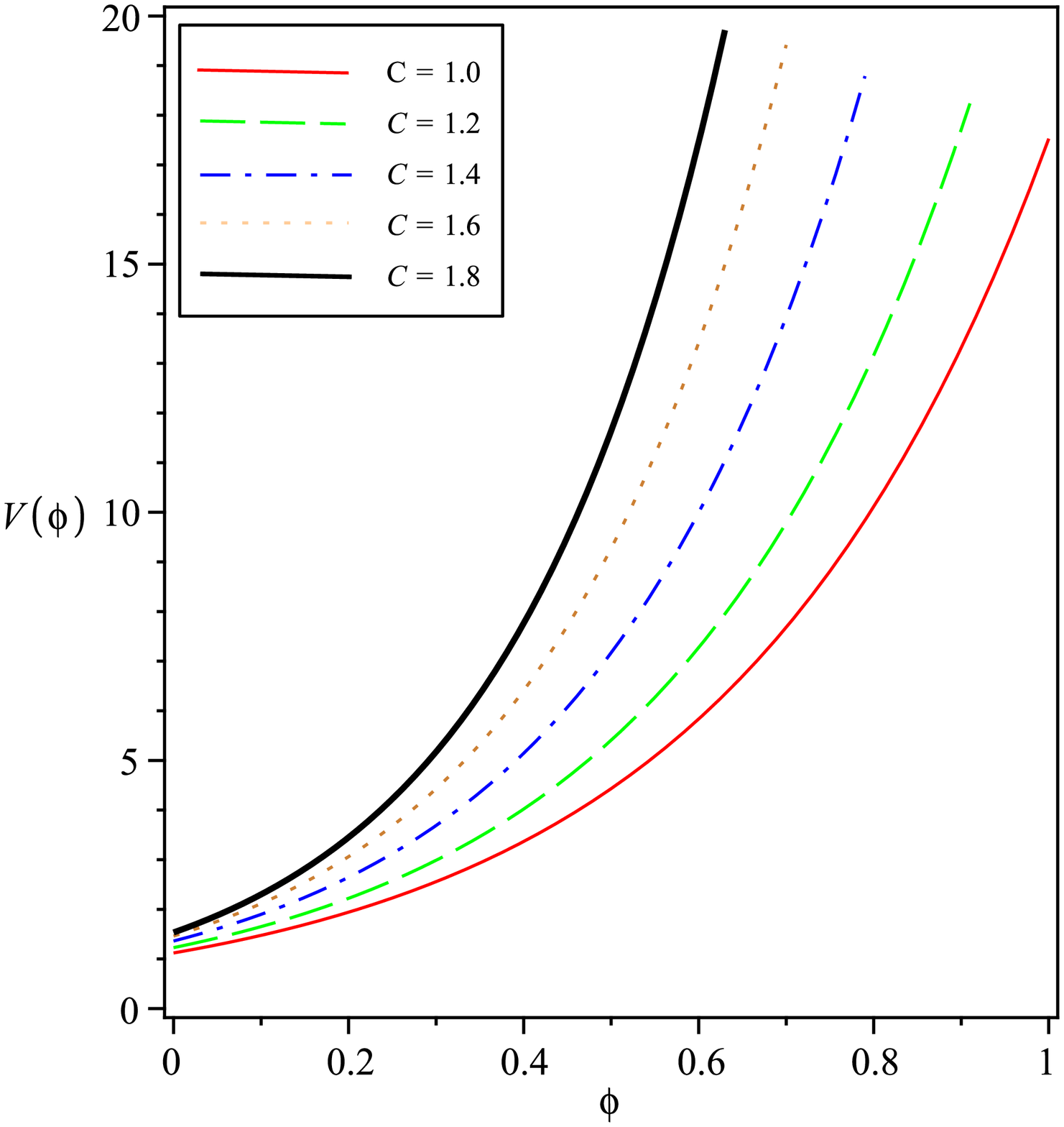, ,width=0.45\linewidth} \\
\end{tabular}
\caption{The evolution of the scalar-field $\phi(a)$ (left) and
the potential $V(\phi)$ (right) for HDE with different parameter
$c$ where $\phi$ is in unit of $m_p$ and $V(\phi)$ in $\rho_{c0}$.
Here we have taken $\Omega_{D0}=0.72$ and $\Omega_{k}=0.01.$  }
\label{fig2}
\end{center}
\end{figure*}
\begin{eqnarray}\label{vphi2}
V(\phi)&=&M^2_pH^2 \Omega_D\left(2+\frac{\sqrt{\Omega_D}}{c}\cos y\right),\\
\dot{\phi}&=&M_pH \left( 2\Omega_D-\frac{2}{c}{\Omega^{3/2}_D}
\cos y\right)^{1/2}.\label{dotphi2}
\end{eqnarray}
Finally we obtain the evolutionary form of the field by
integrating the above equation
\begin{eqnarray}\label{phi}
\phi(a)-\phi(a_0)=M_p \int_{a_0}^{a}{\frac {da}{a}\sqrt{ 2
\Omega_D-\frac{2}{c}{\Omega^{3/2}_D}\cos y}},
\end{eqnarray}
where $a_0$  is the  present value of the scale factor, and
$\Omega_D$ is given by Eq. (\ref{Omegaq3}). Basically, from Eqs.
(\ref{Omegaq3}) and (\ref{phi}) one can derive $\phi=\phi(a)$ and
then combining the result with (\ref{vphi2}) one finds
$V=V(\phi)$. Unfortunately, the analytical form of the potential
in terms of the scalar field cannot be determined due to the
complexity of the equations involved. However, we can obtain it
numerically. For simplicity we take $\Omega_k\simeq0.01$ fixed in
the numerical discussion. The reconstructed quintessence potential
$V(\phi)$ and the evolutionary form of the field are plotted in
Figs. 2, where we have taken $\phi(a_0=1)=0$. A notable point in
this figure is that the reconstructed potentials for different
values of $c$ have a nonzero value at the present time, which can
be interpreted as a cosmological constant behavior of the model
desirable from the perspective of $\Lambda$CDM model.
\section{Quintessence reconstruction of interacting HDE model  \label{IntHDE}}
\begin{figure*}[htp]
\begin{center}
\begin{tabular}{cc}
\epsfig{file=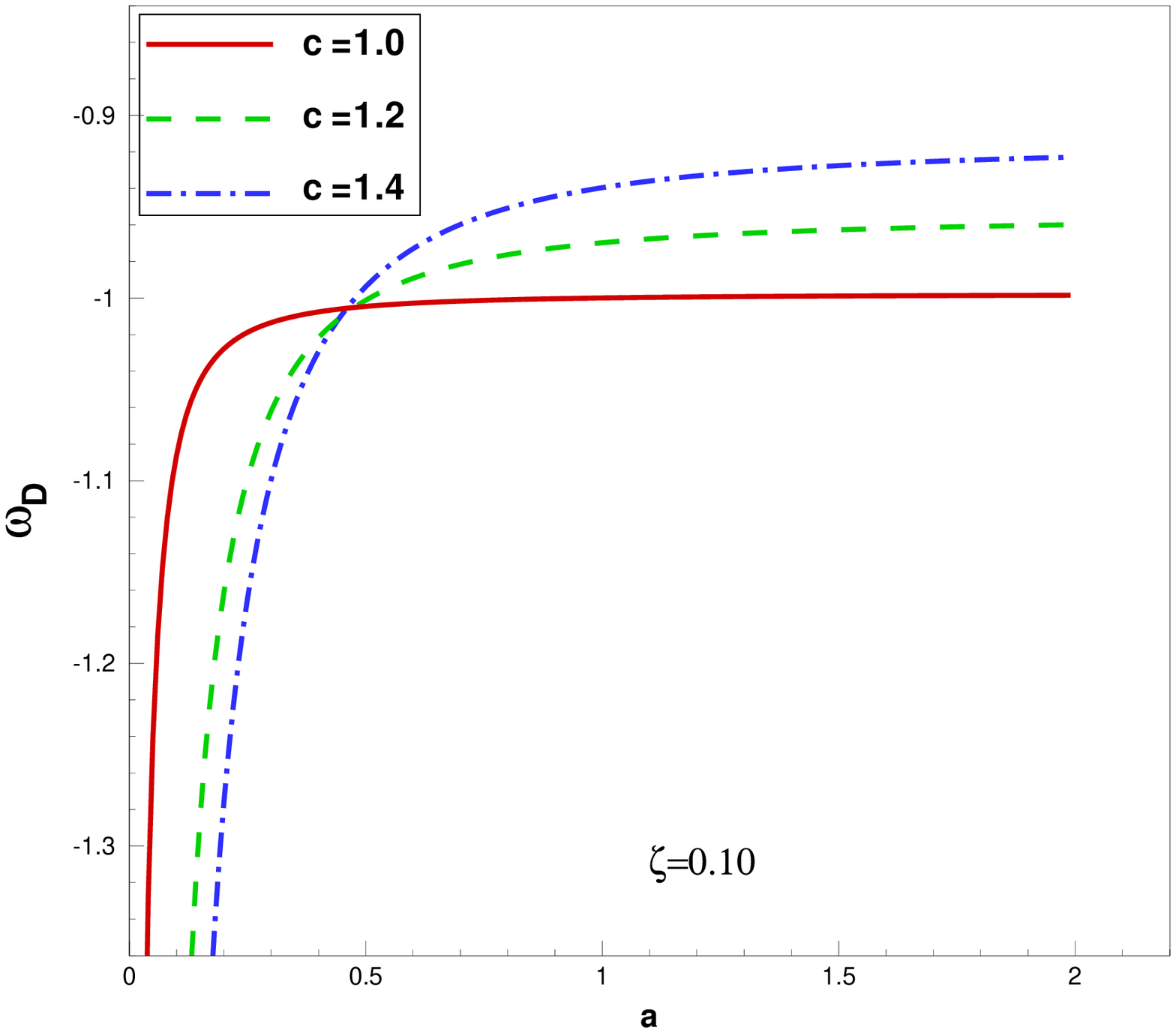, ,width=0.5\linewidth} & \epsfig{file=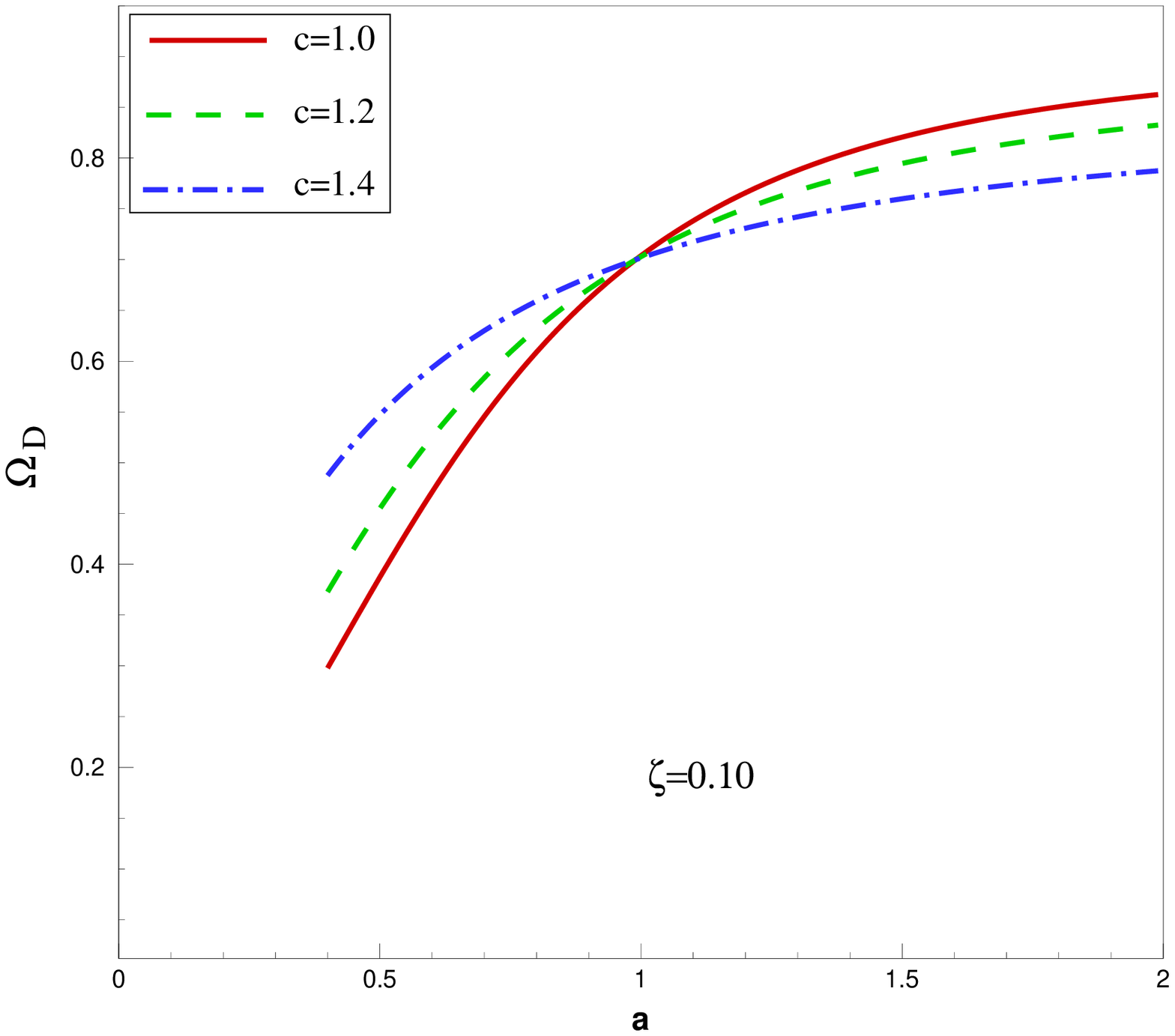, ,width=0.5\linewidth} \\
\end{tabular}
\caption{The evolution of $w_D$ (left) and $\Omega_{D}$ (righ) for
interacting HDE with different parameter $c$. Here we take
$\Omega_{D0}=0.72$ and $\Omega_{k}=0.01.$  } \label{fig3}
\end{center}
\end{figure*}
\begin{figure*}[htp]
\begin{center}
\begin{tabular}{cc}
\epsfig{file=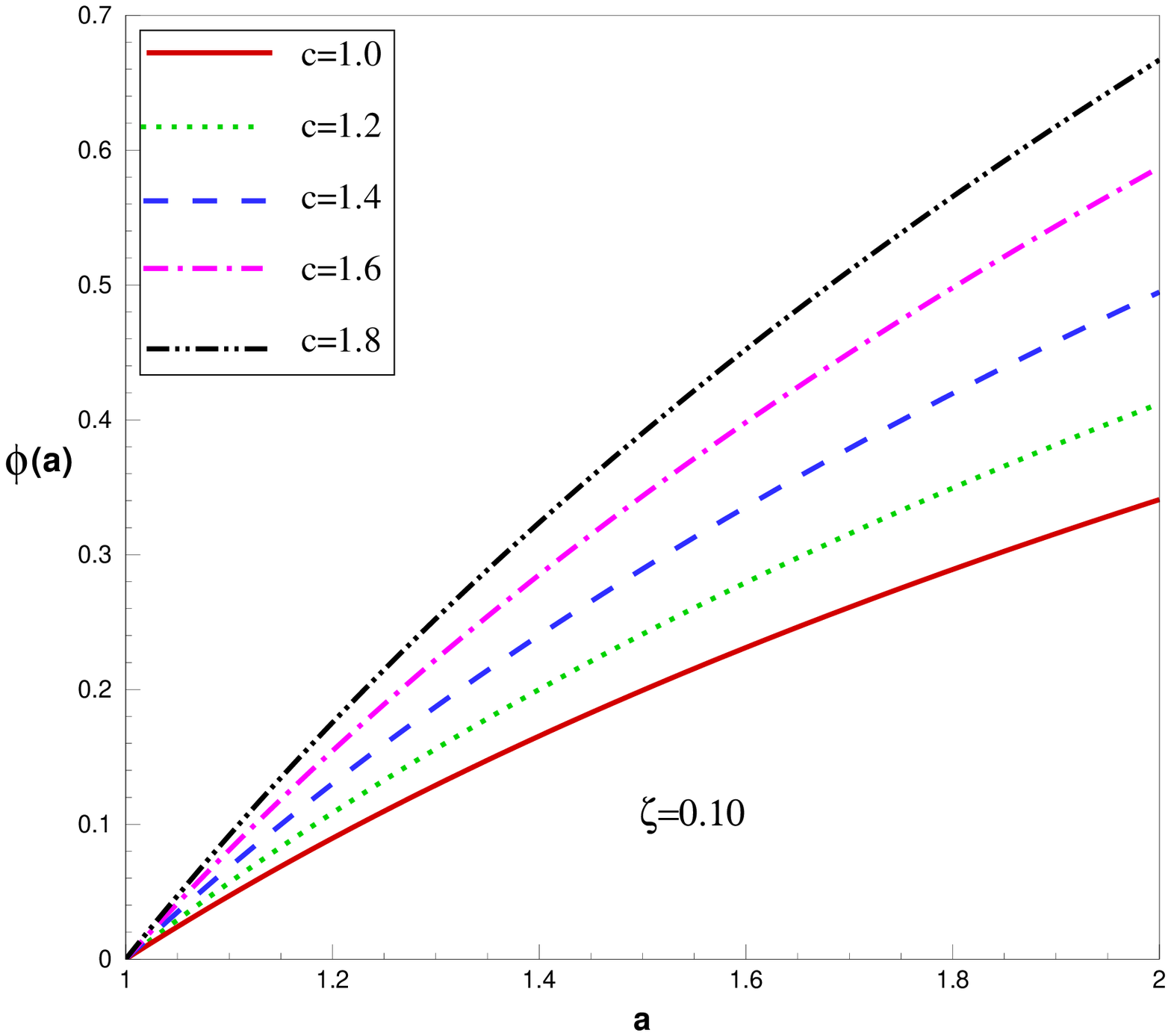, ,width=0.5\linewidth} & \epsfig{file=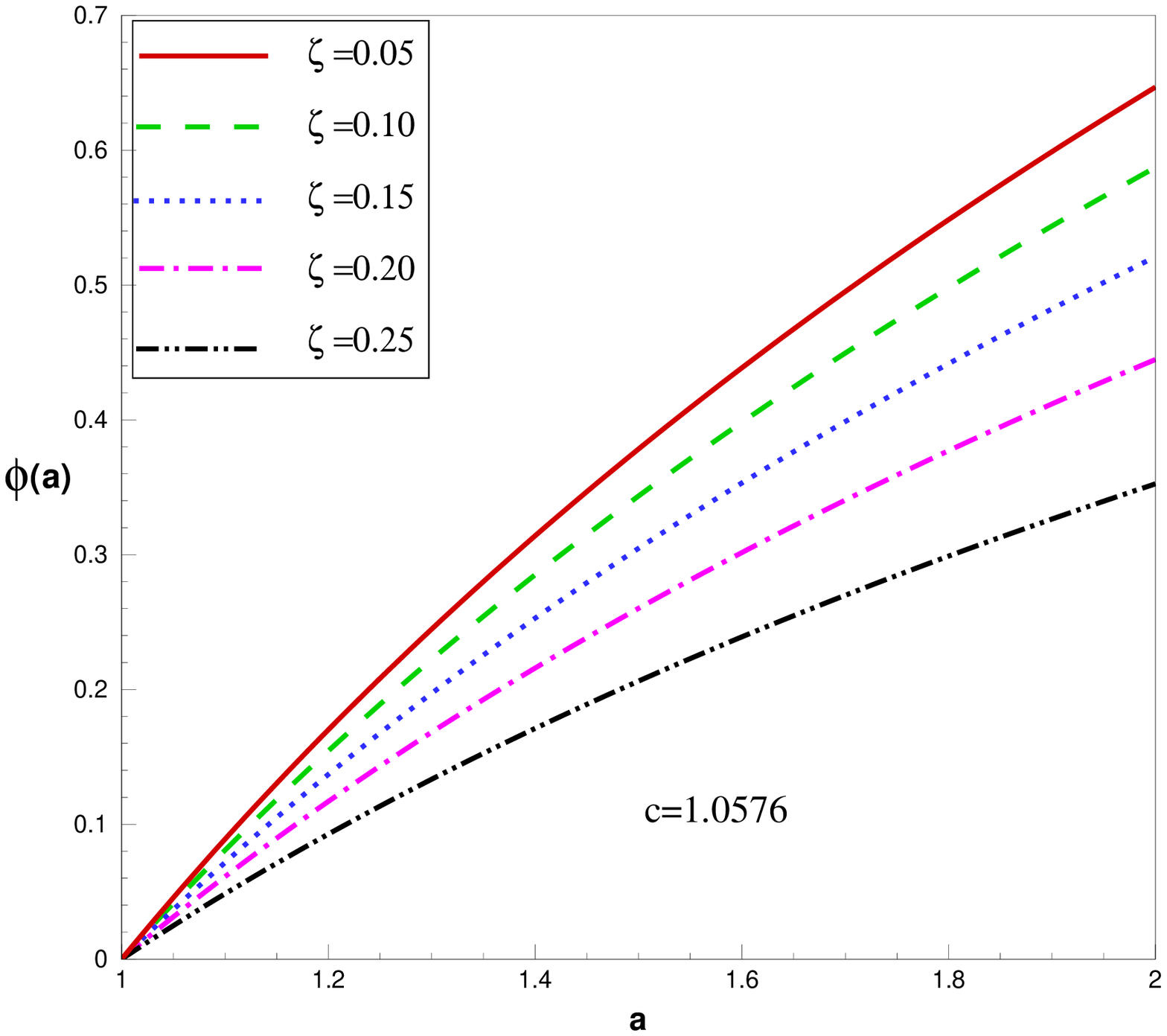, ,width=0.5\linewidth} \\
\end{tabular}
\caption{The revolution of the scalar-field $\phi(a)$ for
interacting HDE with different parameter $c$ (left) and different
parameter $\zeta$ (right), where $\phi$ is in unit of $m_p$ and we
have taken here $\Omega_{m0}=0.28$ and $\Omega_{k}=0.01$.  }
\label{fig4}
\end{center}
\end{figure*}
\begin{figure*}[t!]
\begin{center}
\begin{tabular}{cc}
\epsfig{file=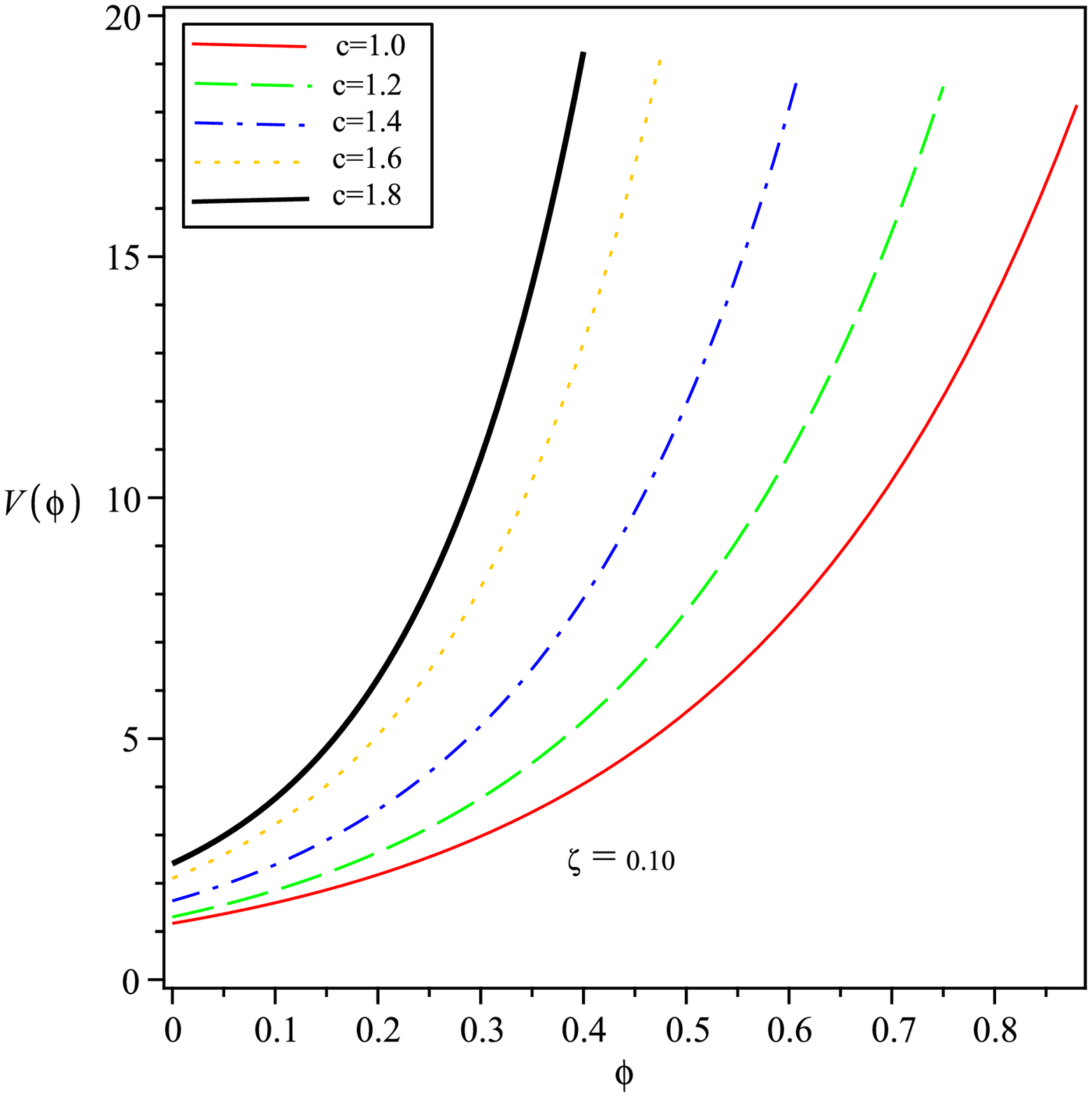, ,width=0.45\linewidth} & \epsfig{file=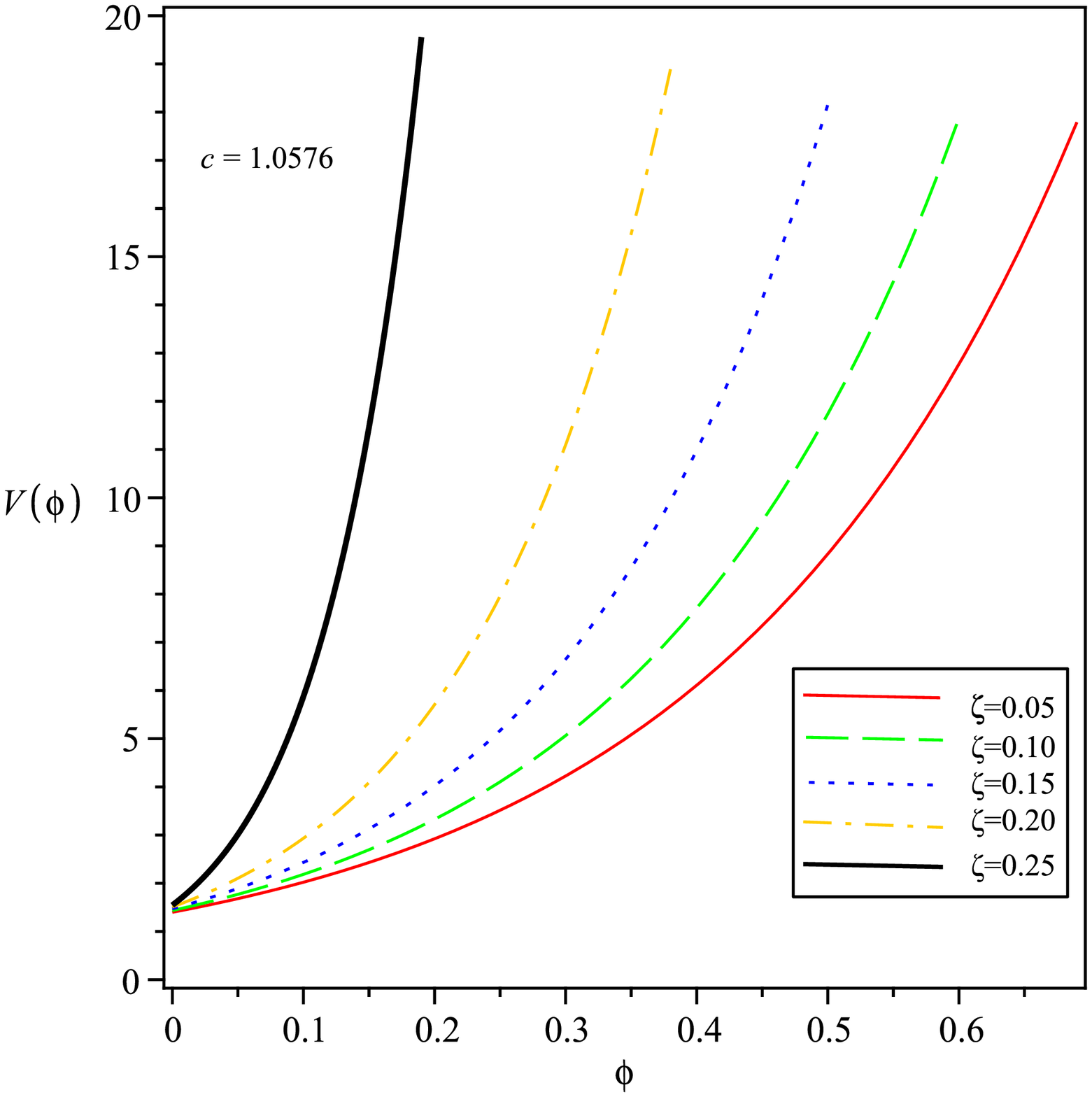, ,width=0.45\linewidth} \\
\end{tabular}
\caption{The reconstruction of the potential $V(\phi)$ for
interacting HDE with different parameter $c$ (left) and different
parameter  $\zeta$ (right), where $\phi$ is in unit of $m_p$ and
$V(\phi)$ in $\rho_{c0}$. We take here $\Omega_{m0}=0.28$ and
$\Omega_{k}=0.01$  } \label{fig5}
\end{center}
\end{figure*}
In this section, we consider the interaction between dark matter
and DE. In this case the continuity equations take the form
\begin{eqnarray}
&&\dot{\rho}_m+3H\rho_m=Q, \label{consm}
\\&& \dot{\rho}_D+3H\rho_D(1+w_D)=-Q.\label{consq}
\end{eqnarray}
where $Q$ denotes the interaction term and can be taken as $Q =3\zeta
H{\rho_D}({1+u})$ with $\zeta$  being a coupling constant and
$u=\rho_m/\rho_D$ is the energy density ratio. Inserting Eq.
(\ref{rhodot}) in conservation law (\ref{consq}), we obtain the
equation of state parameter
\begin{eqnarray}
w_D=-\frac{1}{3}-\frac{2\sqrt{\Omega_D}}{3c}\cos y-\frac{\zeta}{\Omega_D}({1+\Omega_k})\label{wD}.
\end{eqnarray}
Taking the derivative of both side of the Friedman equation
(\ref{Fried}) with respect to the cosmic time, and using Eqs.
(\ref{Fried2}), (\ref{HL}), (\ref{consq}) and (\ref{consm}), it is
easy to show that
\begin{eqnarray}\label{Hdot}
\frac{2\dot{H}}{H^2}=-3-\Omega_k+\Omega_D+\frac{2\Omega^{3/2}_D}{c}+{3{\zeta}({1+\Omega_k})}\label{wD}.
\end{eqnarray}
Substituting this relation into Eq. (\ref{Omegaqn2}), we obtain the
equation of motion of HDE
\begin{eqnarray}\label{Omegaq3}
{\Omega'_D}&=&\Omega_D\left[(1-\Omega_D)\left(1+\frac{2}{c}\sqrt{\Omega_D}\cos
y\right)-{3{\zeta}({1+\Omega_k})}+\Omega_k\right].
\end{eqnarray}
We plot in Fig. 3 the evolutions of  $w_D$ and $\Omega_D$ for
interacting HDE with different parameter $c$. Now we implement a
correspondence between interacting HDE and quintessence scalar
field. In this case we find
\begin{eqnarray}\label{vphi2}
V(\phi)&=&M^2_pH^2 \Omega_D\left(2+{3{\zeta}({1+\Omega_k})}+\frac{\sqrt{\Omega_D}}{c}\cos y\right),\\
\dot{\phi}&=&M_pH \left(2\Omega_D-3{\zeta}({1+\Omega_k})-\frac{2}{c}{\Omega^{3/2}_D}
\cos y\right)^{1/2}.\label{dotphi2}
\end{eqnarray}
Finally, the evolutionary form of the field can be obtained by
integrating the above equation. We obtain
\begin{eqnarray}\label{phi}
\phi(a)-\phi(a_0)=M_p \int_{a_0}^{a}{\frac {da}{a}\sqrt{ 2
\Omega_D-\frac{2}{c}{\Omega^{3/2}_D}\cos y-3{\zeta}({1+\Omega_k})}},
\end{eqnarray}
where $\Omega_D$ is given by Eq. (\ref{Omegaq3}). Again, the
analytical form of the potential in terms of the scalar field
cannot be determined due to the complexity of the equations
involved and we do a numerical discussion. The reconstructed
quintessence potential $V(\phi)$ and the evolutionary form of the
field are plotted in Fig. 4 and 5, where we have taken
$\phi(a_0=1)=0$. For simplicity we take $\Omega_k\simeq0.01$ fixed
in the numerical discussion. In the interacting case there exist a
different manner of evolution for $w_D$. In the pervious section
we found that increasing $c$ leads a faster evolution for $w_D$
toward more negative values while in the interacting case
increasing $c$ cause $w_D$ to evolve toward less negative values
which can predict a slower rate of expansion for the future HDE
dominated universe. Also One can find from Figs. 4 and 5 that the
reconstructed potentials evolve toward a nonzero minima at the
present as mentioned in the noninteracting case.
\section{Model Fitting}\label{fitting}
In this section we will fit the interacting Quintessence  HDE
model, in a flat universe, by using the cosmological data. To
obtain the best fit values of the model parameters, we apply the
maximum likelihood method. In this method the total likelihood
function $\mathcal{L}_{\rm total}=e^{-\chi_{\rm tot}^2/2}$ can be
defined as the product of the separate likelihood functions of
uncorrelated observational data with
\begin{equation}
 \chi^2_{\rm tot}=\chi^2_{\rm SNIa}+\chi^2_{\rm CMB}+\chi^{2}_{\rm BAO}+\chi^2_{\rm gas}\;,\label{totchi1}
\end{equation}
where SNIa stands for type Ia supernovae, CMB for cosmic microwave
background radiation, BAO for baryon acoustic oscillation and {\it
gas} stands for X-ray gas mass fraction data. The details of
obtaining each $\chi$ is discussed in \cite{Alavirad:2013nfy}.
Best fit values of parameters are obtained by minimizing
$\chi_{\rm tot}^2$. In the current paper we will use CMB  data
from seven-year WMAP \cite{Komatsu:2010fb},  type Ia supernovae
data from 557 Union2 \cite{Amanullah:2010vv}, baryon acoustic
oscillation  data from SDSS DR7 \cite{Percival:2009xn}, and the
cluster X-ray gas mass fraction data from the Chandra X-ray
observations \cite{Allen:2007ue}. We apply a Markov chain Monte
Carlo (MCMC) simulation on the parameters of the model by using
the publicly available CosmoMC code \cite{Lewis:2002ah} and
considering the parameter vector $\{\Omega_{\rm b}h^2,\Omega_{\rm
DM}h^2, \zeta\}$.

\begin{figure*}[t!]
\centering
\begin{tabular}{ccccc}
\epsfig{file=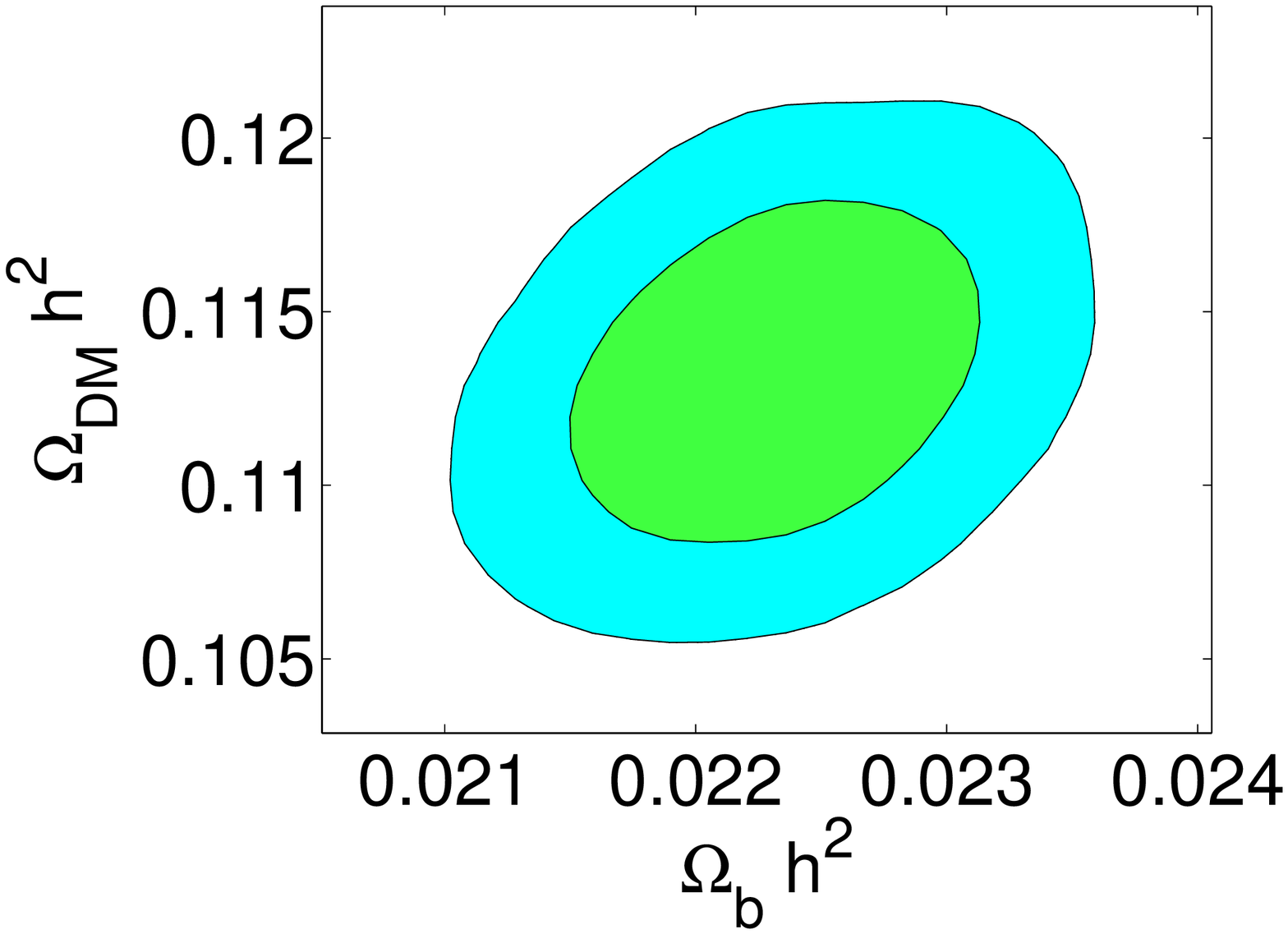,width=0.18\linewidth} & & & &\\
\epsfig{file=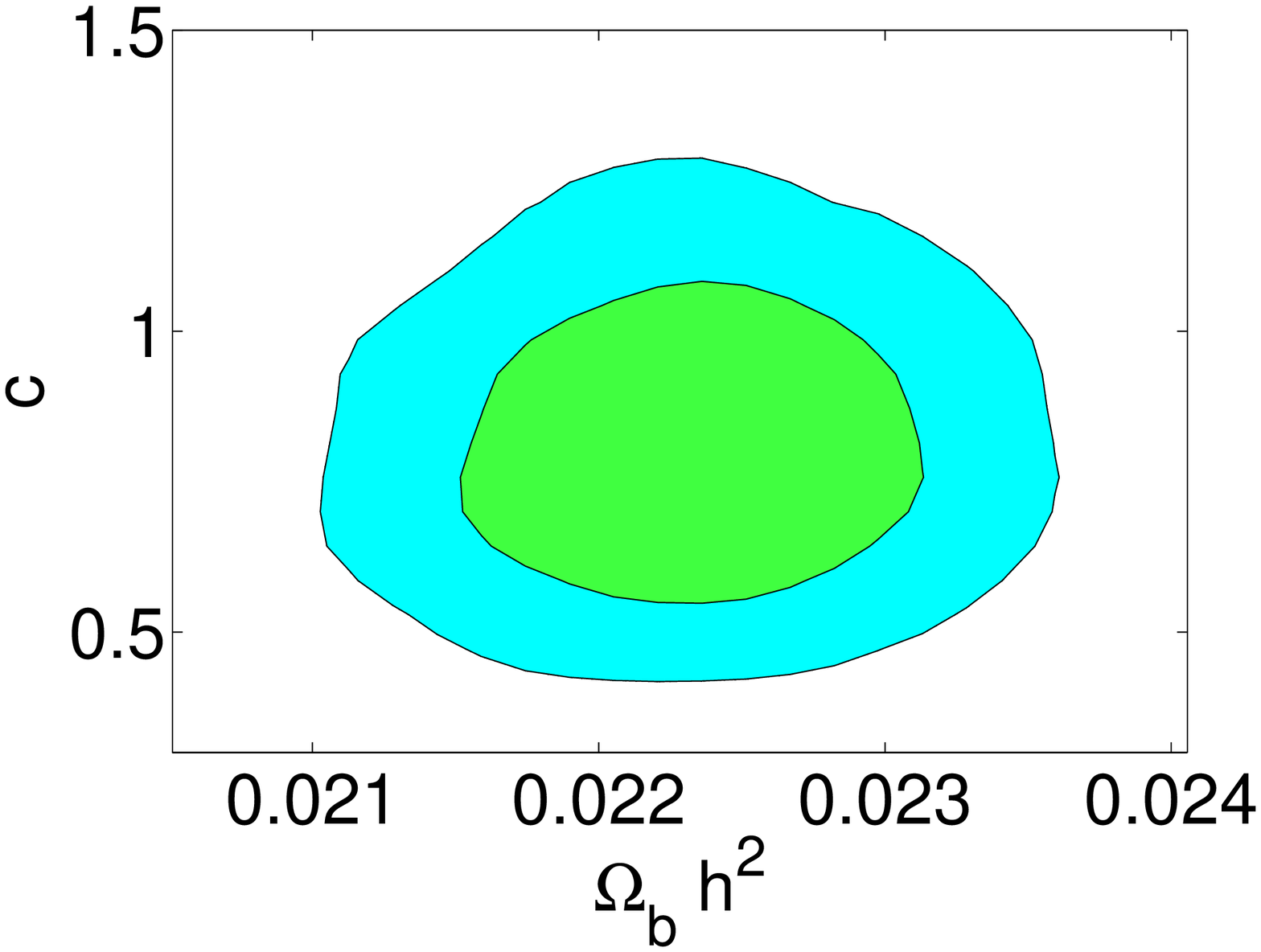,width=0.18\linewidth} & \epsfig{file=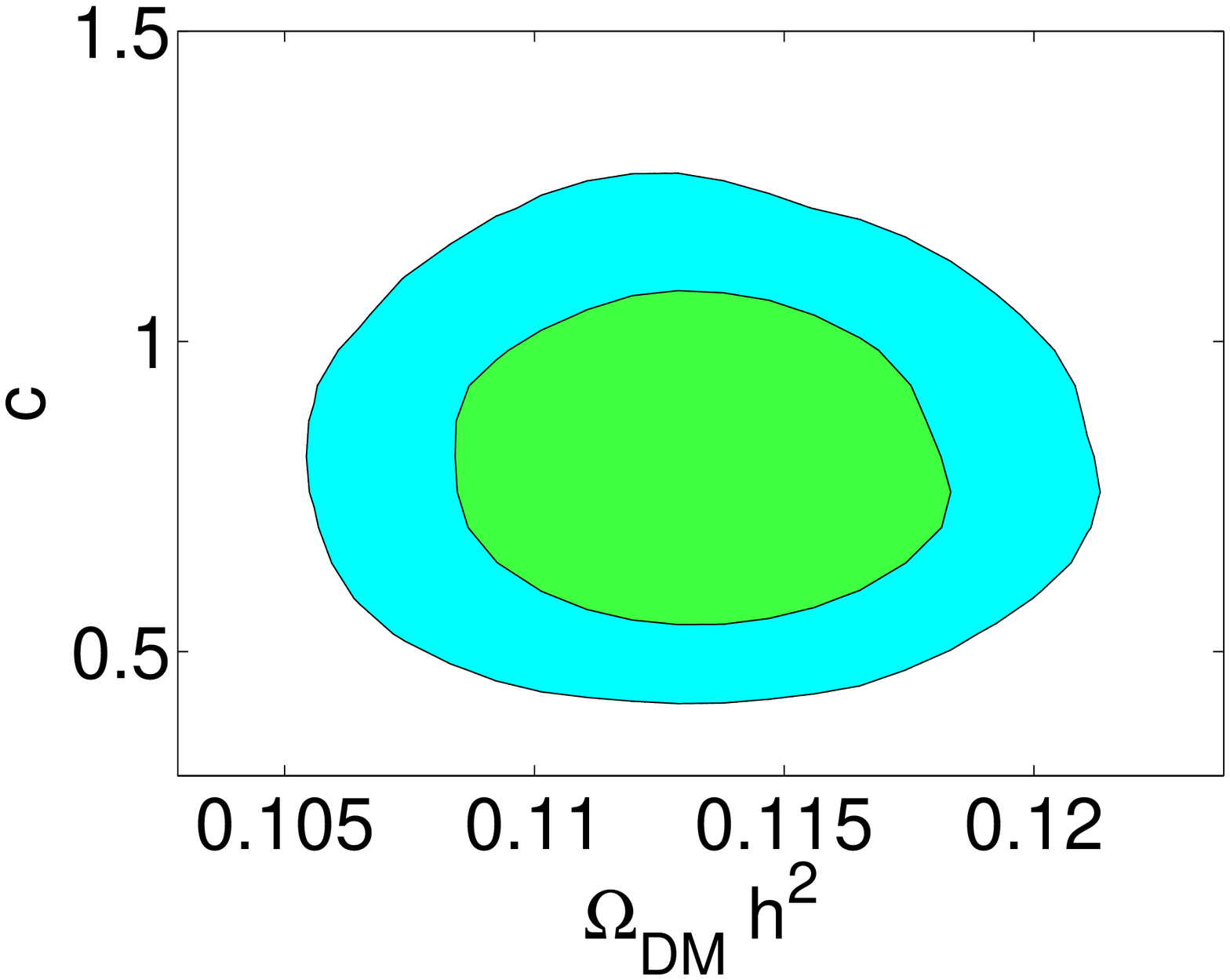,width=0.18\linewidth}& & &\\
\epsfig{file=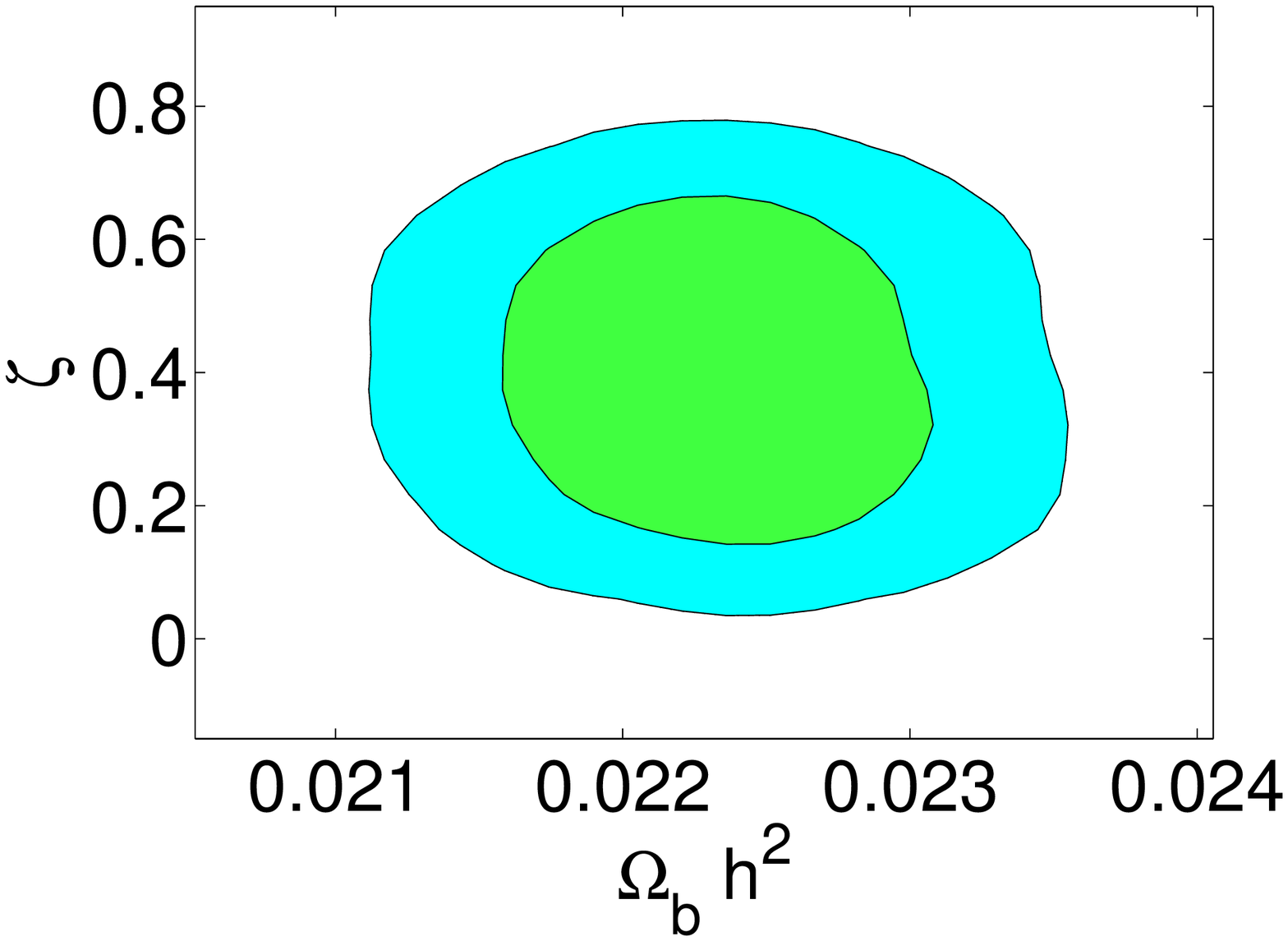,width=0.18\linewidth} &\epsfig{file=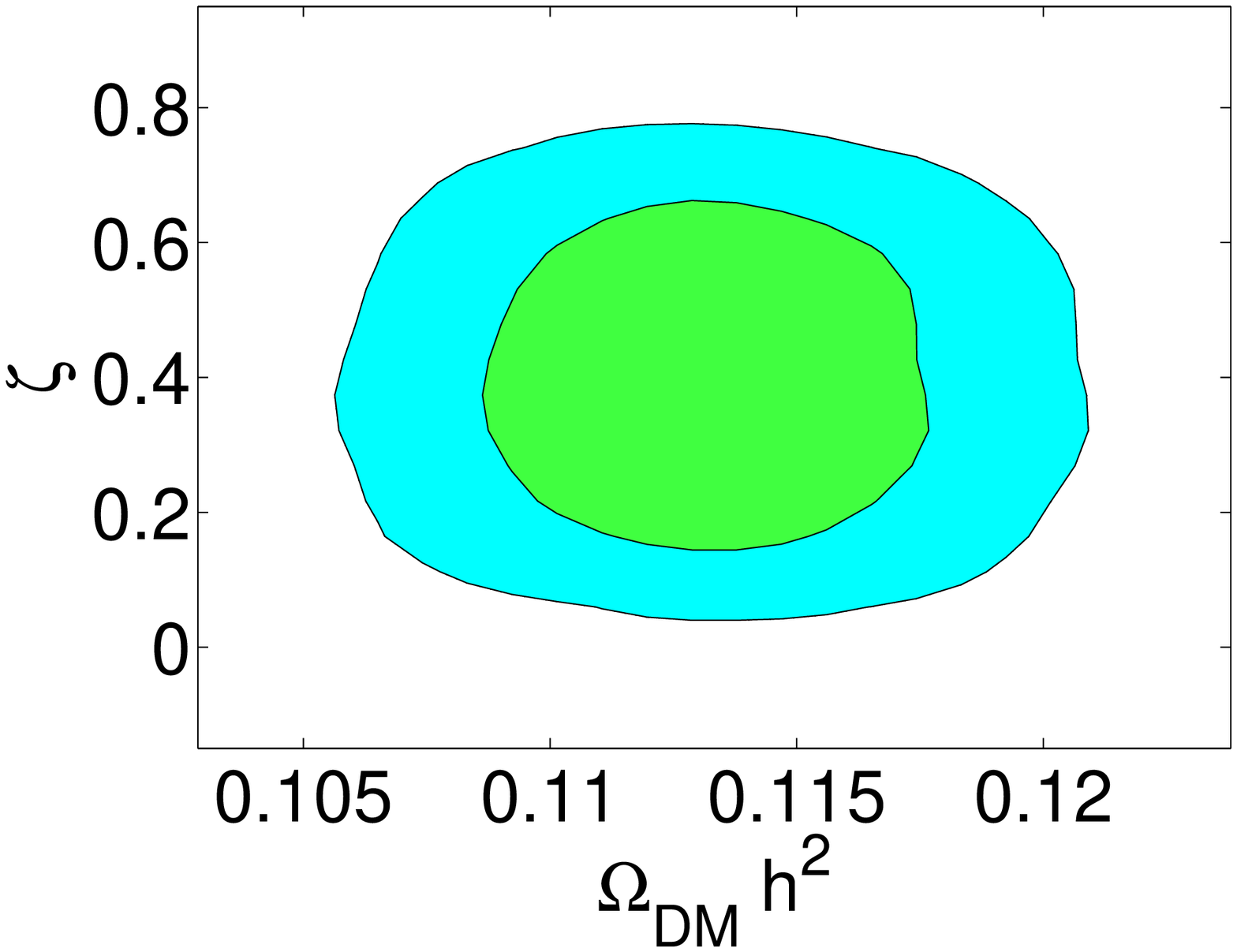,width=0.18\linewidth} & \epsfig{file=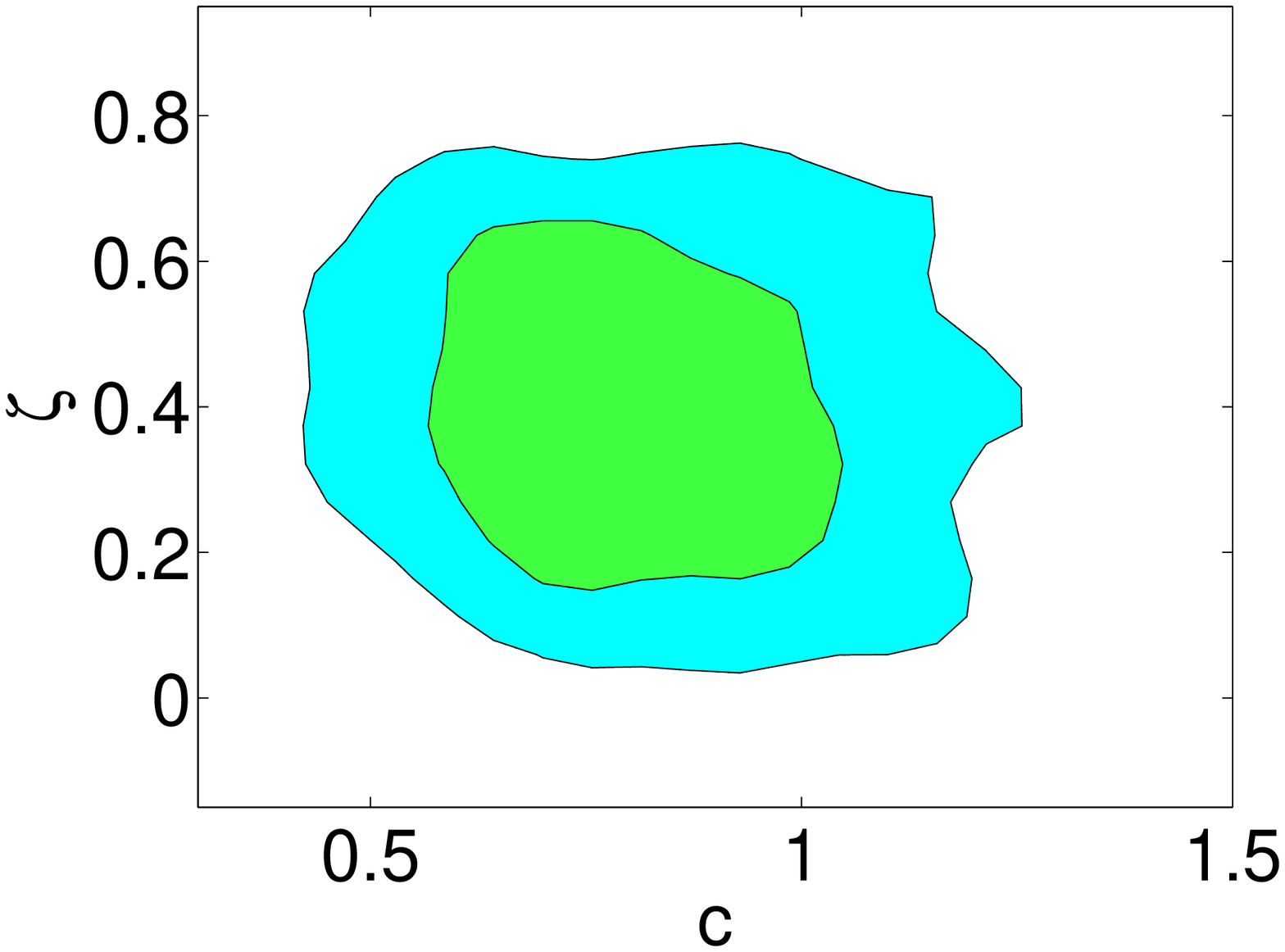,width=0.18\linewidth}& &\\
\epsfig{file=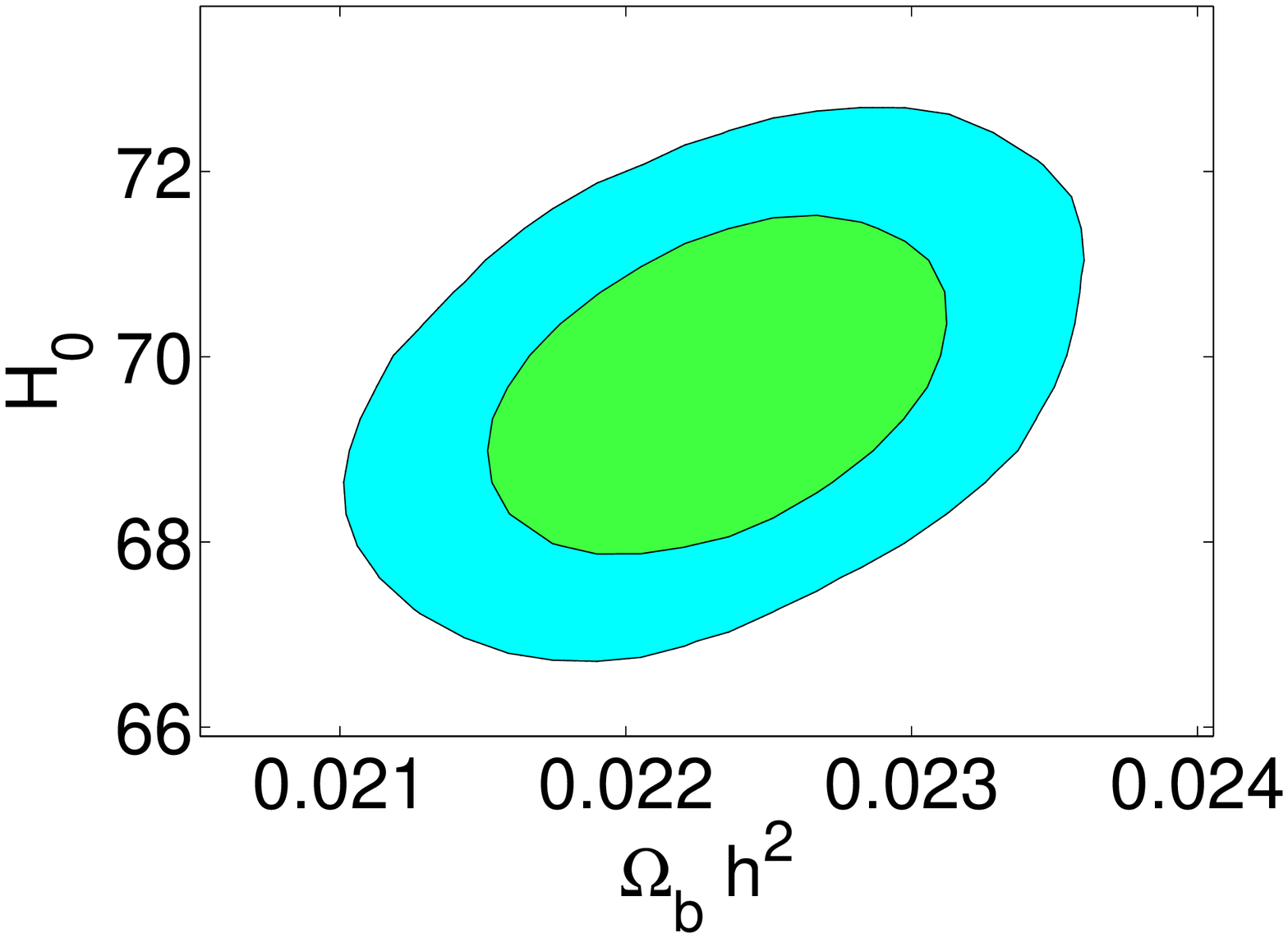,width=0.18\linewidth} &\epsfig{file=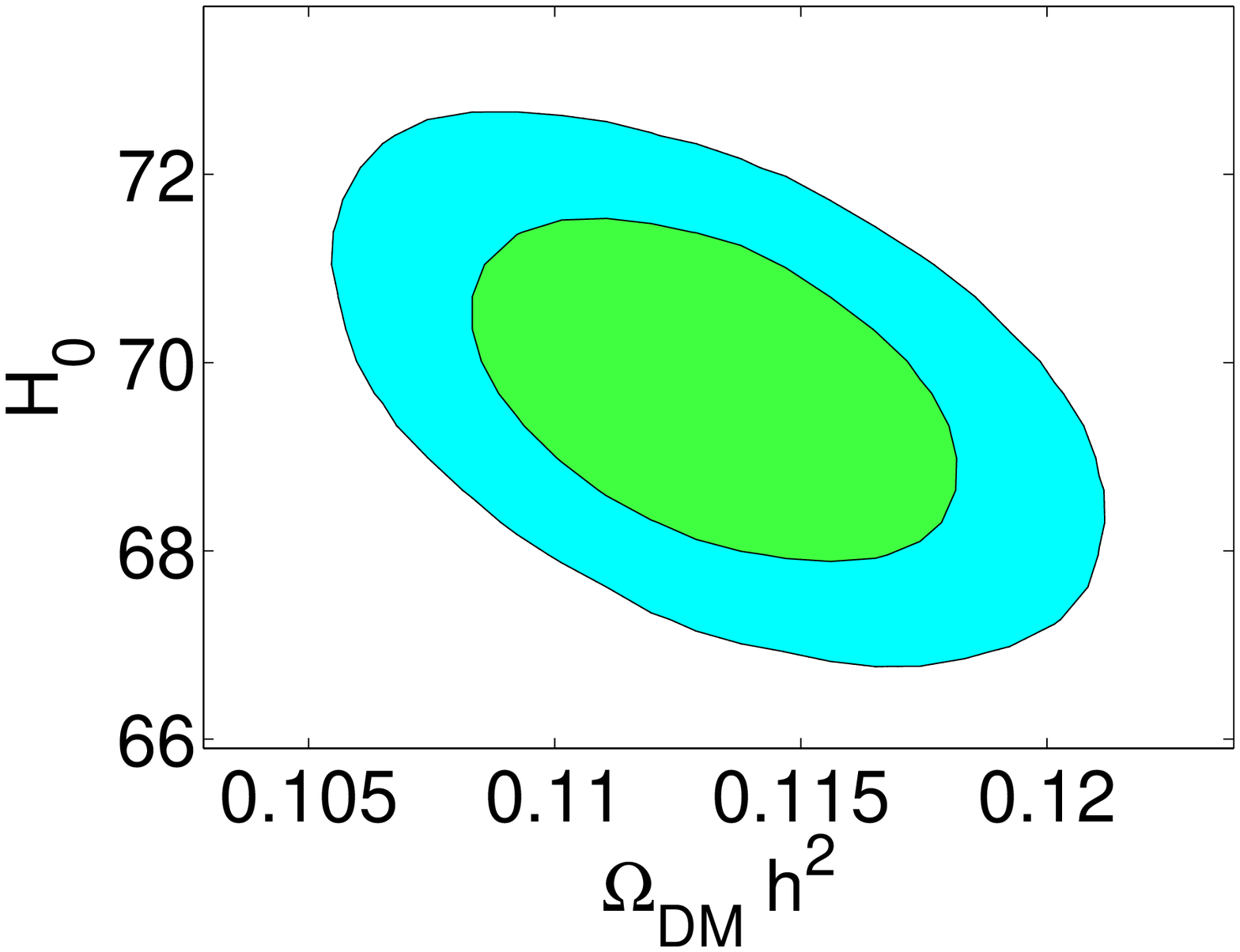,width=0.18\linewidth} & \epsfig{file=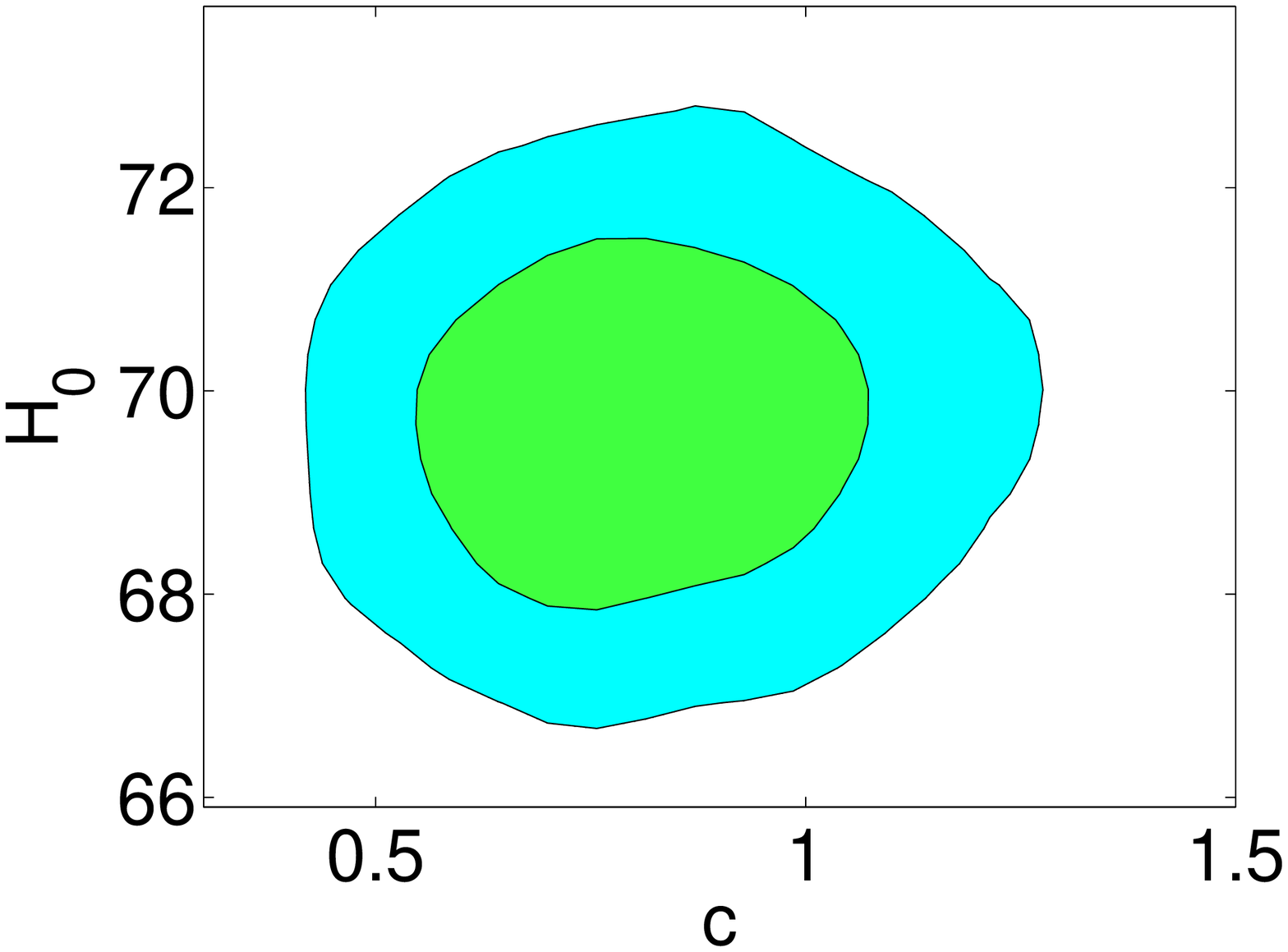,width=0.18\linewidth}& \epsfig{file=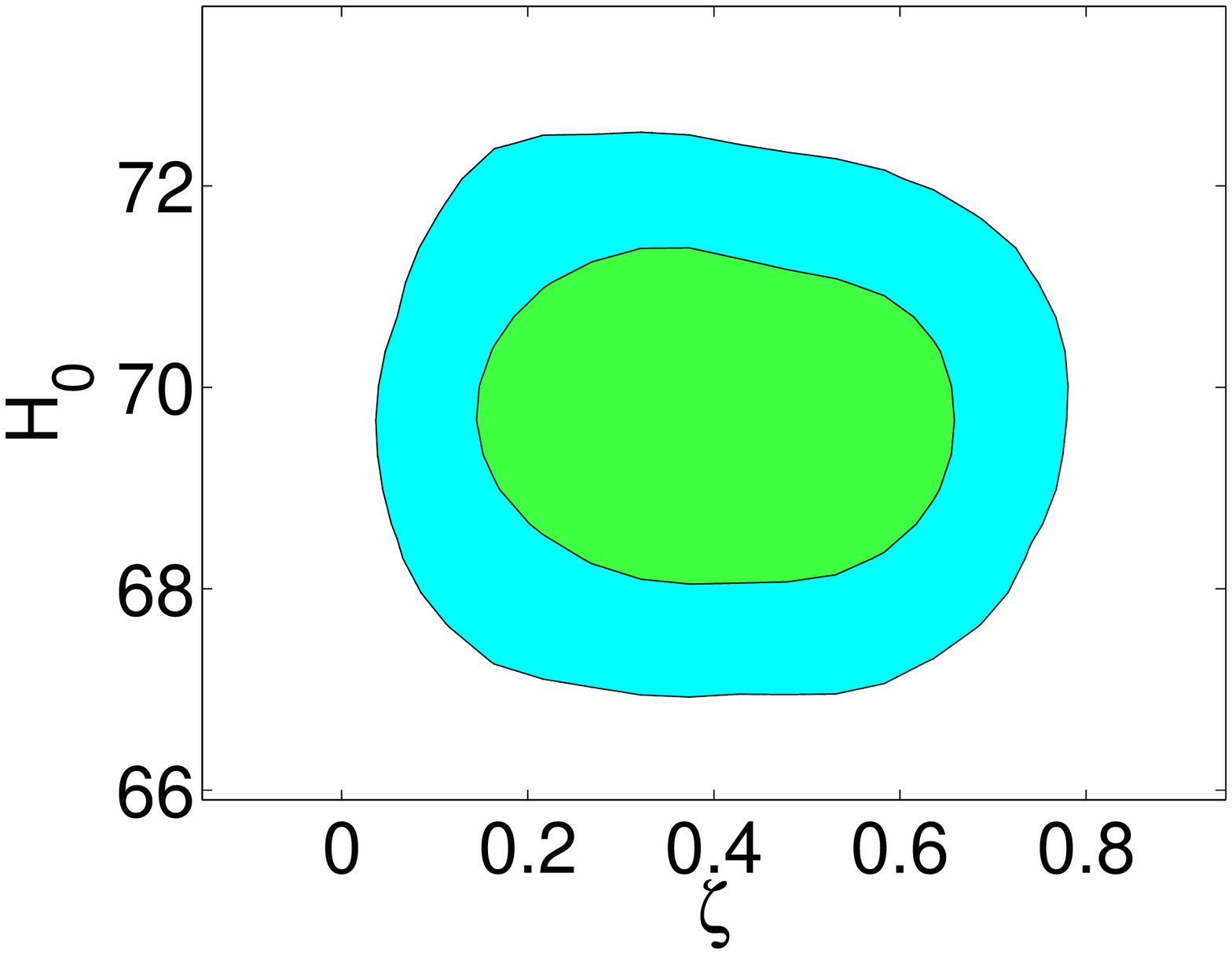,width=0.18\linewidth}&\\
\epsfig{file=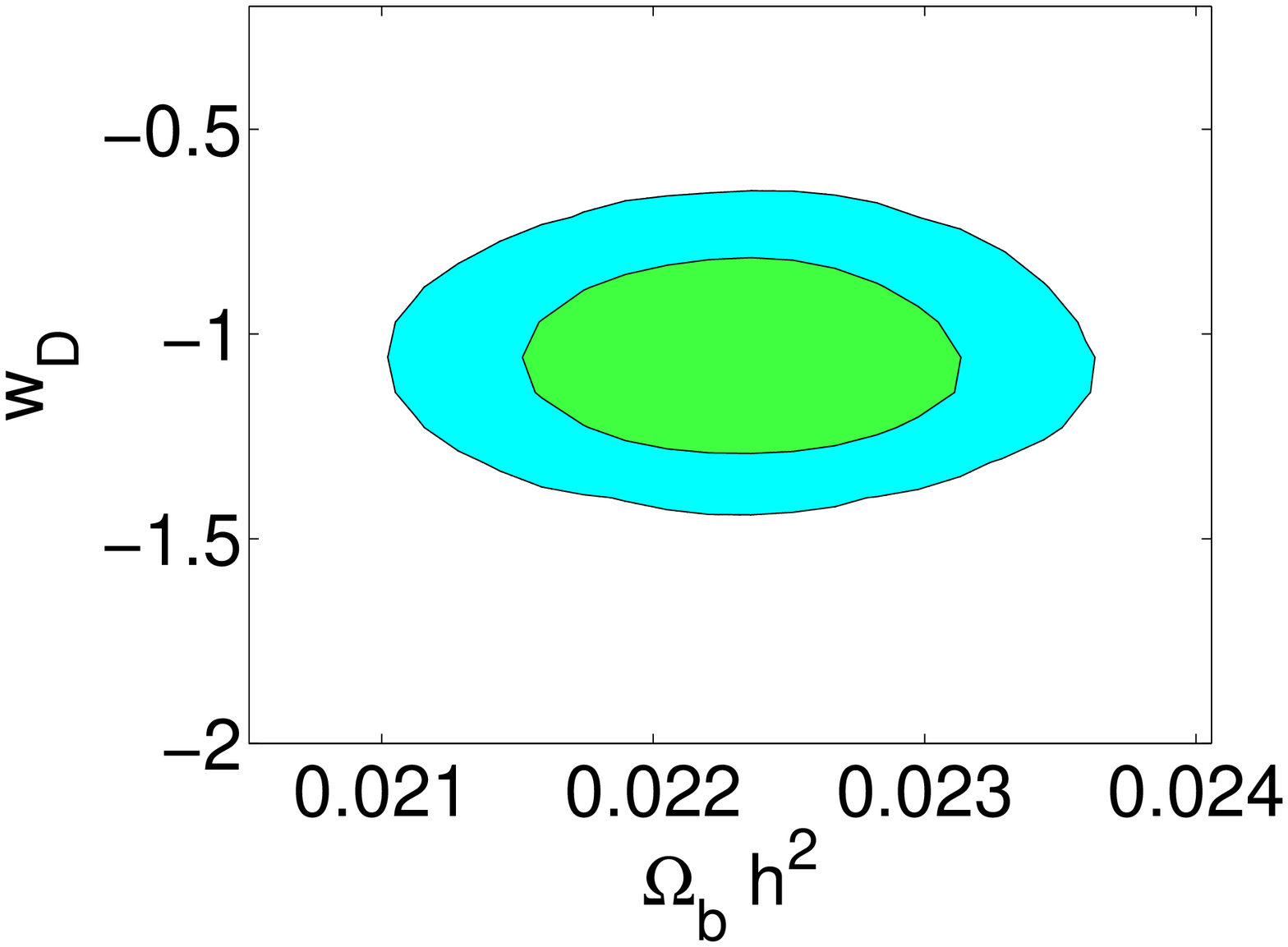,width=0.18\linewidth} &\epsfig{file=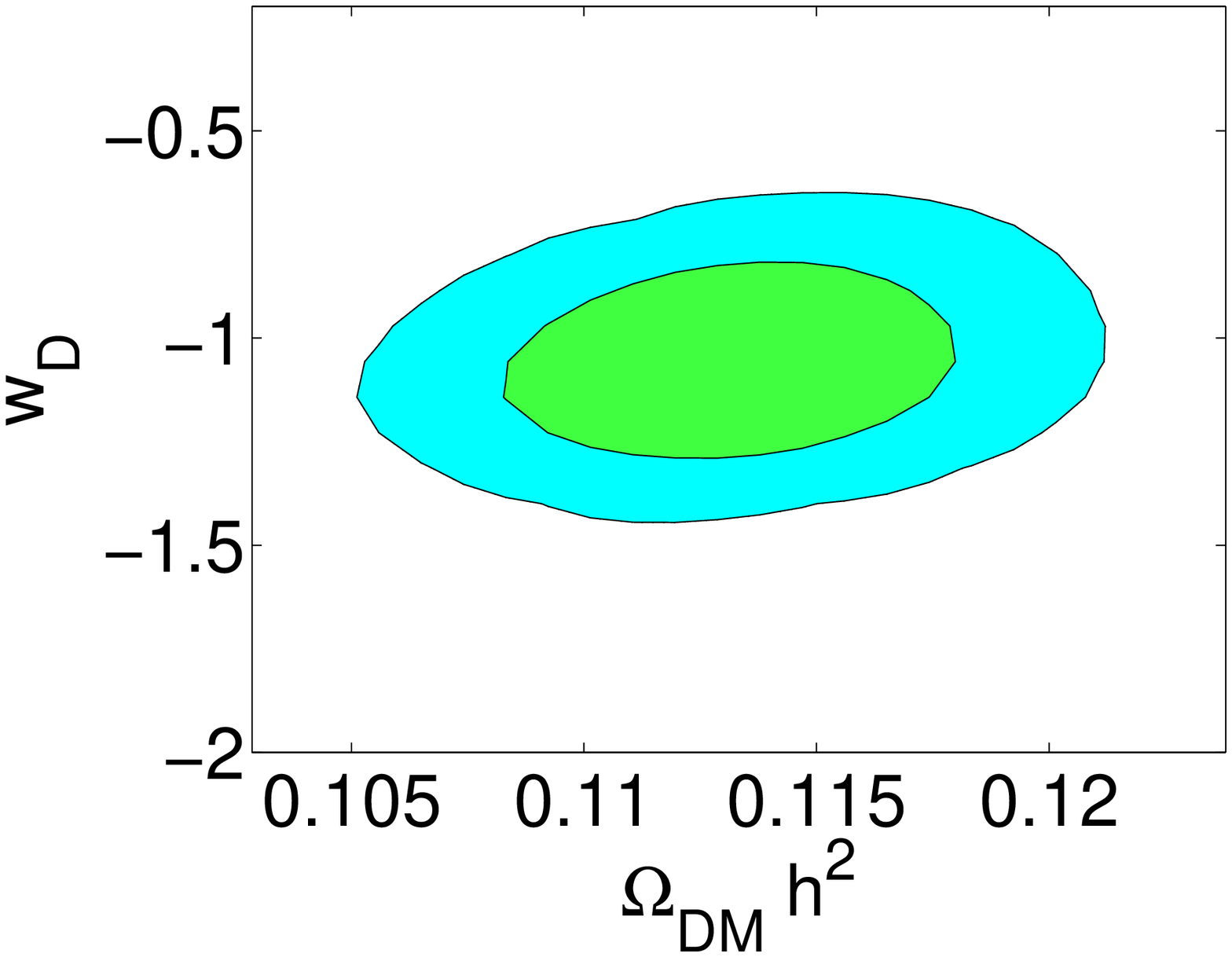,width=0.18\linewidth} & \epsfig{file=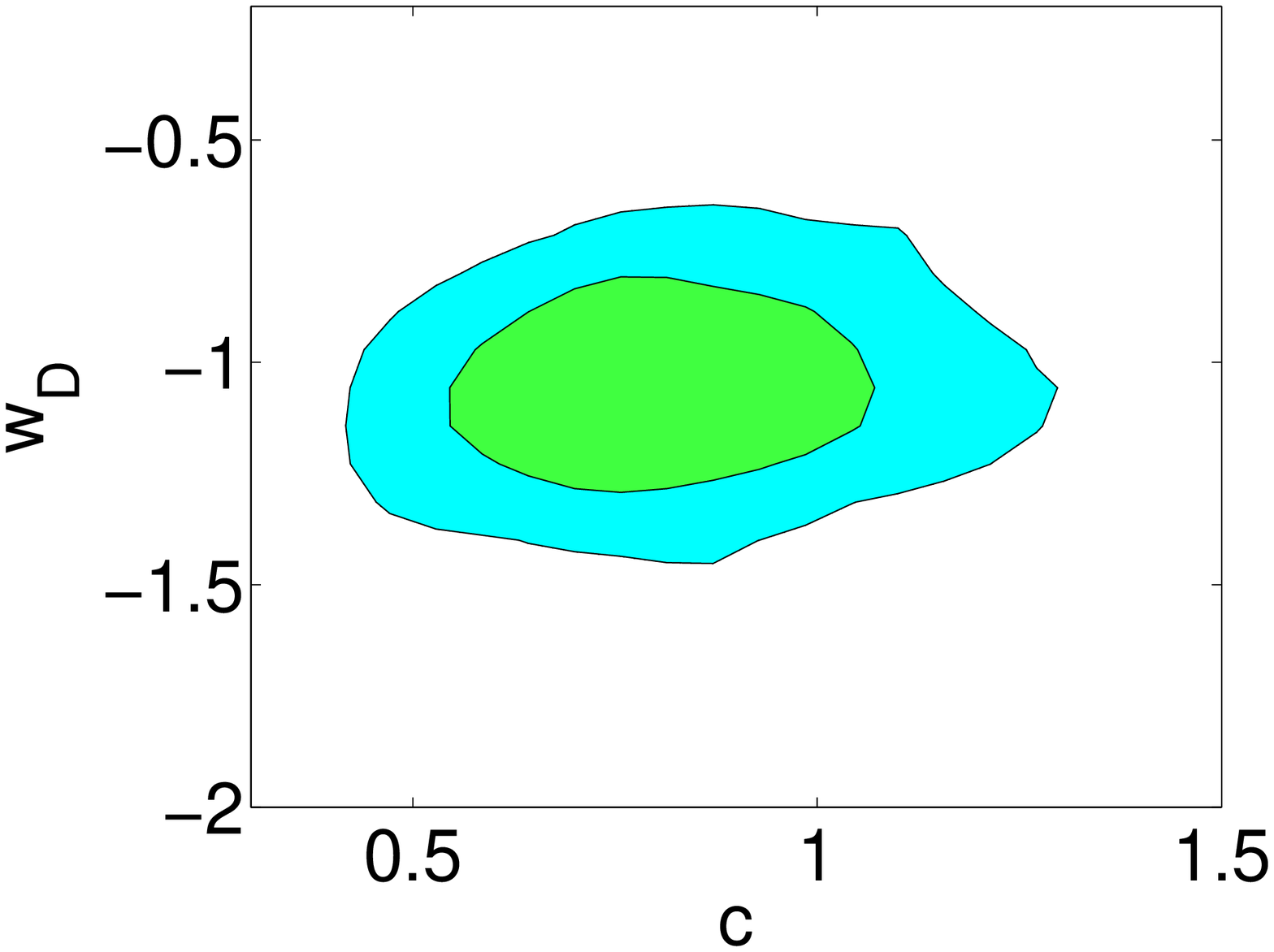,width=0.18\linewidth}& \epsfig{file=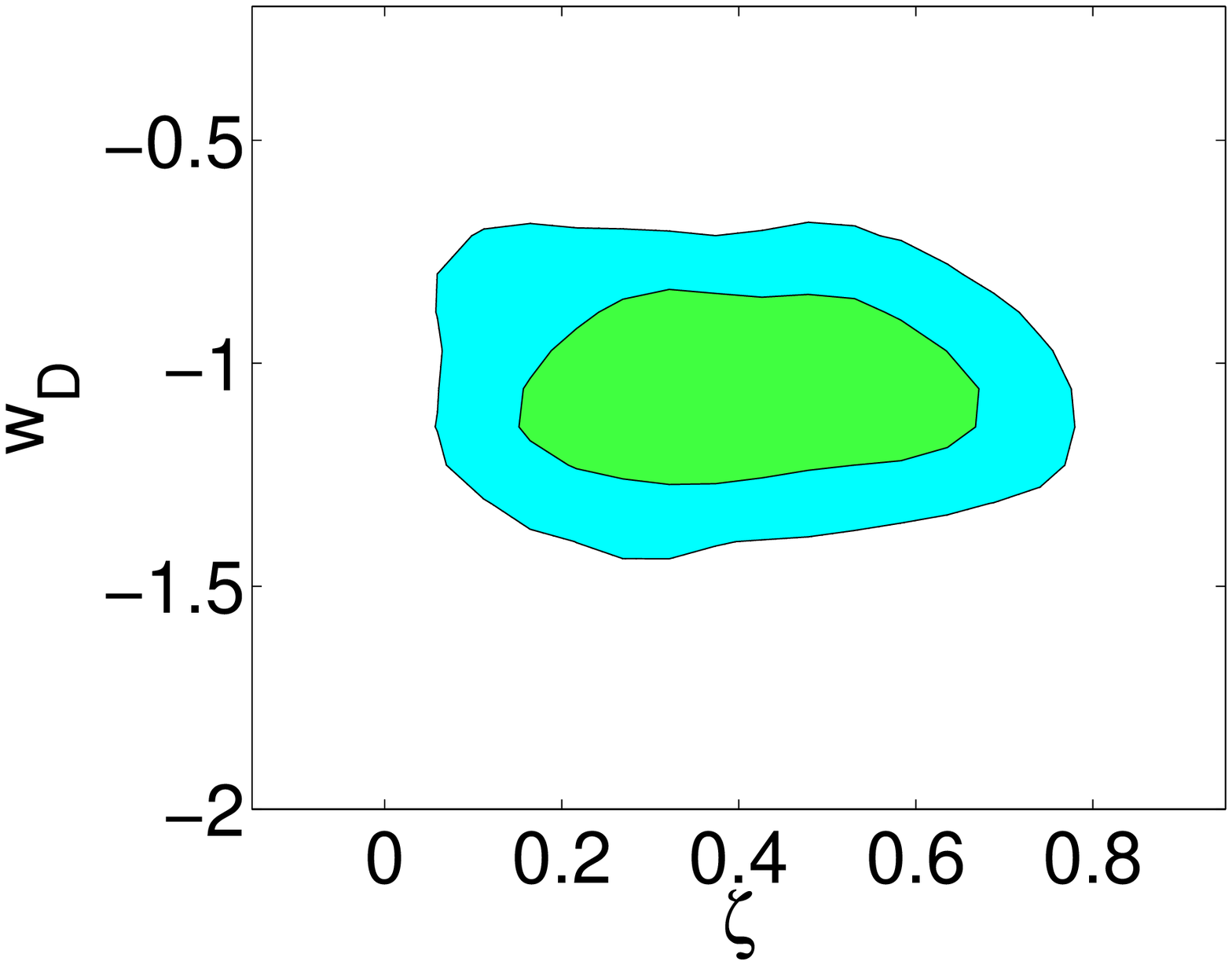,width=0.18\linewidth}& \epsfig{file=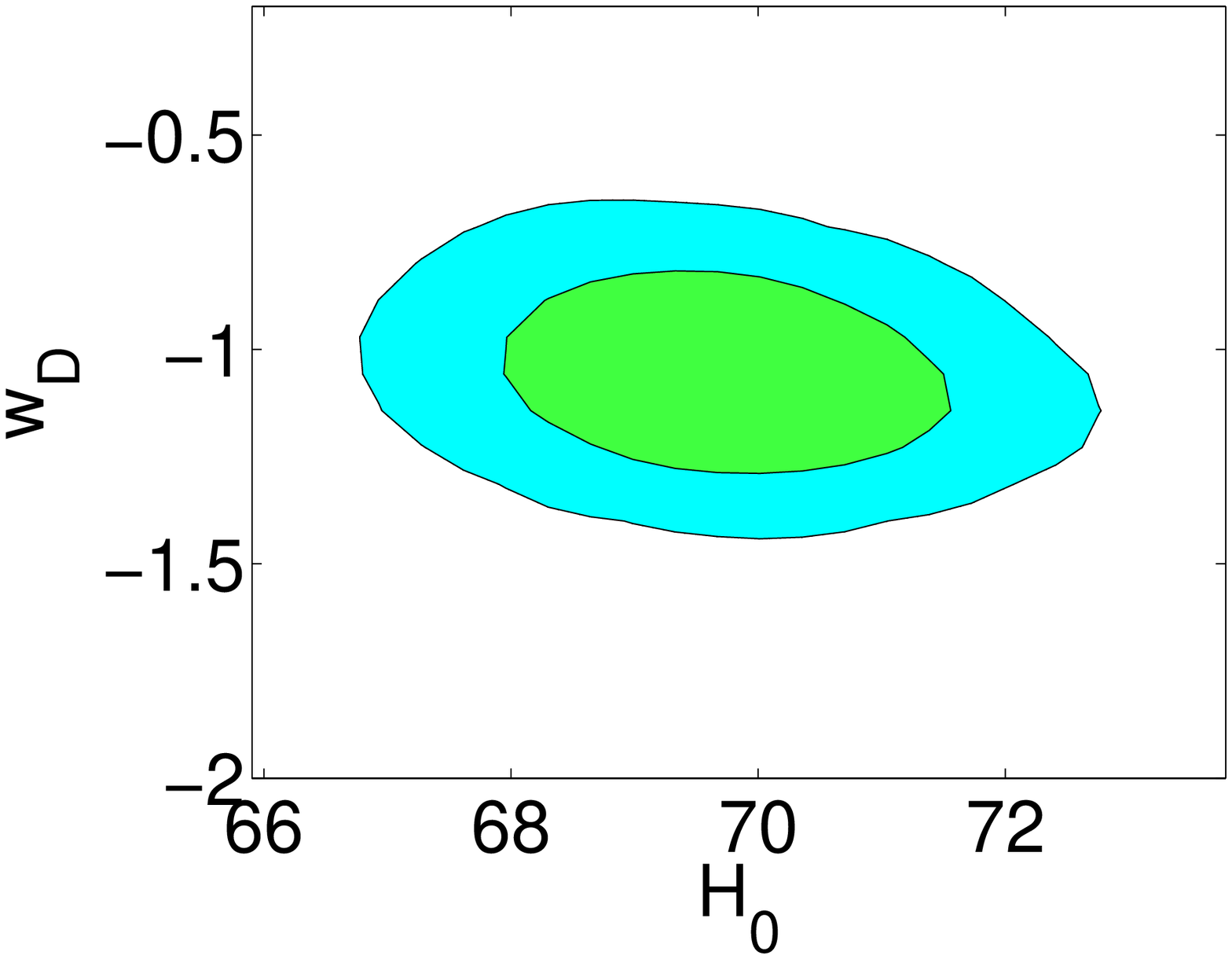,width=0.18\linewidth}\\
\end{tabular}
\caption
{2-dimensional constraint of the cosmological and model parameters contours
in the flat interacting Quintessence  HDE model with $1\sigma$ and $2\sigma$ regions. To produce these plots,
Union2+CMB+BAO+X-ray gas mass fraction data together with the BBN constraints have been used.}\label{fig:MCMC}
\end{figure*}

The MCMC simulation results are summarized in table I and the two
dimensional contours are plotted in figure \ref{fig:MCMC}. From
table I it is clear that the main cosmological parameters
$\Omega_{\rm b}h^2$, $\Omega_{\rm DM}h^2 $, $\Omega_{\rm D}$  are
compatible with the results of the $\Lambda$CDM model
\cite{Ade:2013zuv} as one can see from the third column in table
I. We can see that the best fit value for the dark energy equation
of state crossed the phantom line where
$w_D=-1.249243^{+0.624537+0.636920}_{-0.455913-0.455913}$. In
addition the best fit value of the HDE parameter
$c=1.0576^{+0.3010+0.3052}_{-0.6632-0.6632}$ is compatible  with
the previous numerical analysis works such as
$c=0.91^{+0.21}_{-0.18}$ in \citep{Zhang:2007sh},
$c=0.84^{+0.14}_{-0.12}$ in \citep{Zhang:2013mca} and
$c=0.68^{+0.03}_{-0.02}$ in \citep{Zhai:2011pp}. Here we obtained
a positive best fit value for the interacting parameter in
1$\sigma$ and 2$\sigma$ confidence levels as
$\zeta=0.2433^{+0.6373+0.6373}_{-0.2251-0.2251}$ in spite of we
took negative values in the prior of the parameter $\zeta$ as
well. This positive value suggests only conversion of dark matter
to dark energy. Therefore in this model there is no chance for
converting of DE to dark matter. The interacting parameter in the
HDE model has been constrained by observational data by many
authors although with different parametrization of the interaction
term $Q$ \cite{Chimento:2003iea, Zimdahl:2002zb, Valiviita:2009nu,
Pettorino:2012ts, Salvatelli:2013wra, He:2010im}. In
\cite{He:2010im} the authors have considered the same
parametrization as ours in this paper but they have chosen the
prior on parameter $\zeta$ as $\zeta=[0, 0.02]$ and therefore
obtained the best fit value of parameter $\zeta$ as
$\zeta=0.0006\pm0.0006$.
\begin{table*}[t]\label{tab:MCMC}
\begin{center}
    \begin{tabular}{|c||c|c|}
        \hline
        Parameter     & best fit value& $\Lambda$CDM \\ \hline\hline
        $\Omega_{\rm b}h^2$ & ~   $0.0223^{+0.0016+0.0018}_{-0.001756-0.0019}$   & $0.02214\pm0.00024$       \\ \hline
        $\Omega_{\rm DM}h^2 $& ~    $0.1099^{+0.0126+0.0160}_{-0.0066-0.0090}$  & $0.1187\pm0.0017 $          \\ \hline
        $\Omega_{\rm D}$  & ~      $0.7358^{+0.0304+0.0367}_{-0.0598-0.0700}$    & $0.692\pm0.010 $   \\ \hline
    $c$         & ~       $1.0576^{+0.3010+0.3052}_{-0.6632-0.6632}$    & \ldots  \\ \hline
    $\zeta$           &~ $0.2433^{+0.6373+0.6373}_{-0.2251-0.2251}$      & \ldots                     \\ \hline
        $H_{0}$         & ~      $70.7745^{+3.0981+3.6730}_{-4.6435-5.1342}$   & $67.80\pm0.77$     \\ \hline
        $w_{D}$     &~ $-1.2492^{+0.6245+0.6369}_{-0.4559-0.4559}$  & -1 \\\hline
        \end{tabular}
\caption{The best fit values of the cosmological and model parameters in the  interacting Quintessence  HDE model in a flat universe with $1\sigma$ and $2\sigma$ regions.
Here CMB, SNIa and BAO and X-ray mass gas fraction data together with the BBN constraints have been used. For comparison,
the results for the $\Lambda$CDM model from the Planck data are presented as well \cite{Ade:2013zuv}.}
\end{center}
\end{table*}

\section{Conclusions\label{CONC}}
Adopting the viewpoint that the quintessence scalar field model of
DE is the effective underlying theory of DE, we are able to
establish a connection between the quintessence scalar field and
interacting HDE scenario. This connection allows us to reconstruct
the quintessence scalar field model according to the evolutionary
behavior of interacting holographic energy density. We have
reconstructed the potential as well as the dynamics of the
quintessence scalar field which describe the quintessence
cosmology. Unfortunately, the analytical form of the potential in
terms of the scalar field cannot be determined due to the
complexity of the equations involved. However, we have plotted
their evolution numerically. A close look at these figures shows
several notable points. In the noninteracting case, we found that
increasing $c$ leads to a faster evolution for $w_D$ toward more
negative values, while in the interacting case, increasing $c$
cause $w_D$ to evolve toward less negative values which can
predict a slower rate of expansion for the future HDE dominated
universe. Also the evolutionary behavior of the potential,
$V(\phi)$, revealed that in both interacting/noninteracting cases
the potential evolves a non zero value at the present time
implying a cosmological constant behavior of the model in this
epoch of its evolution.

By constraining the cosmological parameters of the quintessence
HDE model  in a flat universe, we found that the best fit values
of the main cosmological parameters $\Omega_{\rm b}h^2$,
$\Omega_{\rm DM}h^2 $, $\Omega_{\rm D}$ are in agreement with the
$\Lambda$CDM model as one can see from table I. The best fit
values of the HDE parameter $c$ and interacting parameter $\zeta$
are compatible with the results of the previous constraining works
on the HDE in the presence of interaction between DE and dark
matter. Moreover, according to our data fitting our model can
cross the phantom line in $1\sigma$ confidence level in the
present time of the Universe expansion.

\acknowledgments{We thank the referee for constructive comments
which helped us to improve the paper significantly. A. Sheykhi
thanks the Research Council of Shiraz University. The work of A.
Sheykhi has been supported financially by Research Institute for
Astronomy \& Astrophysics of Maragha (RIAAM), Iran.}

\end{document}